\documentclass[preprint,3p]{elsarticle}
\usepackage{graphicx,t1enc}
\usepackage{amssymb,amsmath}
\usepackage[colorlinks=true,allcolors=blue]{hyperref}
\usepackage{xcolor}
\usepackage{lipsum}
\usepackage{caption}
\usepackage{subcaption}
\usepackage{ulem}

\journal{Chaos, Solitons \& Fractals}

\begin{document}

\begin{frontmatter}
\title{Non-local interaction effects in models of interacting populations}

\author[sal1,sal2]{Mario I. Simoy}
\ead{misimoy@exa.unicen.edu.ar}

\author[cab1,cab2,cab3]{Marcelo N. Kuperman}
\ead{kuperman@cab.cnea.gov.ar}

\address[sal1]{Instituto de Investigaciones en Energ\'{\i}a No Convencional (CONICET - UNSa) , Salta,  Argentina.}
\address[sal2]{Instituto Multidisciplinario sobre Ecosistemas y Desarrollo Sustentable (UNCPBA - CICPBA),  Tandil,  Argentina.}

\address[cab1]{FiEstIn, Gerencia de F\'isica, Centro At\'omico Bariloche.}
\address[cab2]{Instituto Balseiro, Universidad Nacional de Cuyo.}
\address[cab3]{Consejo Nacional de Investigaciones Cient\'ificas y T\'ecnicas, 8400 S. C. de Bariloche, Argentina.}

\begin{abstract}
We consider a couple of models for the dynamics of the populations of two interacting species, inspired by Lotka-Volterra's classical equations. The novelty of this work is that the interaction terms are non local and the interaction  occurs within a bounded range.
These terms include the competitive intraspecific interaction among individuals and the interspecific terms for which we consider two cases: Competition and predation.
The results show that not only the non-locality induces spatial structures but also allows for the survival of the species when due to predation or the competitive exclusion extinction was expected, and even promotes spatio-temporal patterns not linked to eventual temporal oscillations in the local case. 
In this work we also explore some interesting details about the behavior of the population dynamics that shows spatial patterns that interfere in a way that leads to non-trivial results.
\end{abstract}


\end{frontmatter}

\section{Introduction}

The population dynamics of interacting species has been extensively and successfully modeled by means of reaction-diffusion equations, such as the well-known Lotka-Volterra's \cite{murray89}, where the reaction terms describe the interactions between individuals of the species involved and the diffusive terms, when present,  explains the simplest transport process. Typically, most models using these equations assume only local interaction terms that might be a good representation of the real scenario when these interactions occur within a tightly confined neighborhood of an individual. However, when broader interaction radius and home ranges are considered, there might be a spatially extended influence not only due to  the environment profile but also because of  the extent of intraspecific and interspecific interactions. Those features require non-local descriptions to be properly taken into account. When placing these models in a biological context, we must first ask ourselves if considering non-local terms is sensible. The assumption that the interactions are local is too strong and many times gratuitous, especially when individuals can move and communicate with each other or when the home range is wide enough to consider the actions of each individual only locally.

Interestingly, the inclusion of non-local  terms, besides helping to address the effects mentioned above adequately, predicts the existence of non-trivial population dynamics with emerging spatial patterns in the distribution of populations of a nature different from that associated with the Turing mechanism. These effects were first reported in \cite{fkk1,fkk2} in the context of a single species and where the non-local term was linked to intraspecific competition. In these works, the authors showed the formation of patterns induced by non-locality in the distribution of a population whose dynamics is described by the FKKP equation. There are several other examples where non-local effects have been considered in single species population \cite{hernan04,clerc05,genie06,cunha11} or in competitive interacting species \cite{maruvka06,hany20,han20b,maciel21}.

The existence of spatial patterns could be especially relevant when seeking an answer to the ubiquitous violation of the principle of competitive exclusion. According to this principle, two species or more that compete for the same resource cannot coexist if environmental factors remain constant \cite{hard60}; only  the  best competitor will survive. However, a violation of this principle has been observed in various ecosystems where a large number of competing species coexist \cite{lack45,mcart58,hutch61,kalmy11}

Many mechanisms have been proposed to account for these discrepancies and to try to bound the validity of the principle. Some attempts point to considering a dynamical environment \cite{chess00,amara03,mac18}, while others look for some shortcuts in the competition due to diet differentiation \cite{kart15} or more complex patterns of interaction \cite{aya71,grill17}.

However, the answer might be much simpler. The inclusion of the non-local effect, as will be shown,  proves to be a simple and sensible mechanism that can easily predict scenarios of coexistence when otherwise it would be forbidden. There are previous works that also consider non-local terms  (see for example \cite{segal13, merchant11, sun21, pal19, tian17}). The underlying models in most of these works consider integrodifferential equations, where the kernel in the integral accounts for  the non-locality. For example,  in \cite{fkk1}, where  a single species $u(x,t)$ was studied, the departure equation  was

\begin{eqnarray}  \label{influeq}
\frac{\partial u\left( \vec x,t\right) }{\partial t} & = & D\nabla^2 u\left(
\vec x,t\right)+a\,u(\vec x,t)  \\
& & -b\,u(\vec x,t) \int_{\Omega} u(\vec y,t) f_\omega ( \vec x,  \vec y).
\nonumber
\end{eqnarray}
where the kernel $f_{\omega}(\vec x,\vec y)$ is a distribution, characterized by a range $\omega$, and normalized in the domain $\Omega$. 
The simplest forms for the kernel are a Gaussian and a square distribution. The square is simply a normalized constant function within a bounded interval and vanishing outside it. In fact, it has been shown that one of the most relevant characteristics of this kernel to give rise to a stable spatial structure is, more than its precise shape, its boundedness \cite{fkk1,fkk2}.

While the appearance of spatial patterns due to the inclusion of nonlinear terms is not a novel result, in this work we will explore more thoroughly some interesting details about the behavior of the population dynamics that shows spatial patterns that interfere in a way that leads to unreported non-trivial results. Besides, here we will also show the emergence of spatio-temporal patterns not observed before. We will also go beyond a competitive interaction and explore a predator-prey system that again reveals nontrivial effects due to the non-locality of the interactions.

In summary, the goal of this paper is twofold. On the one hand, we aim to study the inclusion of non-local aspects in traditional models of interaction between species and analyze the results within an ecological framework. We want to show that the non-local terms are responsible for non-trivial behaviors such as the promotion of coexistence and survival in situations where in the absence of non-locality it is not possible. On the other hand, we want to explore solutions with space-time structures, with direct ecological implications but also relevant in a broader aspect.

In the following sections, we will address two systems considering two interacting species. First, we will analyze a competitive interaction and then a predator-prey system. As will be noted, a plethora of new features emerge due to the inclusion of non-locality and all of them  will be discussed. In what follows we will consider that non-local effects are due to the range of interaction between individuals while the environment is homogeneous and isotropic.

\section{The nonlocal term}
As mentioned before, we will study a set of two populations involved in two types of interaction: Competition and predation. Each case will be associated with a specific pair of integro-differential equations with a common feature, the addition of non-local terms.
These terms account for the fact that a given individual not only interacts with those located in the same place but also with others enclosed in a bounded region of a certain size. 
The strength of the interaction can be weighted by a function of the distance between individuals or can be the same for any individual within the defined range of interaction. In this work, and for the sake of simplicity, we will consider this last option, as done before in \cite{fkk1,fkk2}

After the previous considerations, a typical non-local term will be associated with an integral over the interaction domain, weighting the number of individuals at a certain distance $d$ from the focal subject with an influence function $K(d)$ that in our one-dimensional example will be a constant value $\kappa=1/\omega$ such that $\omega$ is the interaction length.
A typical non-local term will look as 
\begin{equation}
\int_\Omega z(y,t)K_z(|x-y|)dy 
\end{equation} 
with 
\begin{equation}
K_z(x) = \left\{ \begin{array}{l} 1/2\omega_z, \,\,  \mbox{if  } |x| \leq \omega_z \mbox{ and } \omega_z\neq 0\\ \\ \delta(x), \,\,  \mbox{if  } \omega_z=0\\ \\  0,\,\, \mbox{otherwise }  \\
\end{array} \right.
\label{eq:kernel}
\end{equation} 
and $x$ denoting the spatial variable and $\delta(x)$ is Dirac's Delta.
 
\section{Competitive interaction}
\subsection{The model}

First, we will study the competitive interaction between two species $x$ and $y$. Both species feed on the same resource, though the use each of them makes of this resource is not necessarily the same. A simple set of equations to represent such a scenario is \cite{murray89}

\begin{eqnarray}
\frac{dx}{dt}& = & r_x x \left( 1 - \frac{x}{k_x} - b_{xy} \frac{y}{k_x} \right), \nonumber \\
\\
\frac{dy}{dt}& = & r_y y \left( 1 - \frac{y}{k_y} - b_{yx} \frac{U}{k_y} \right), \nonumber 
\end{eqnarray}
where the growth rate of each species is given by $r_x$ and $r_y$ respectively, and the carrying capacity of the environment is  $k_x$ and $k_y$ for $x$ and $y$, respectively. The competition strength is given by  $b_{xy}$ and $b_{yx}$.

A proper adimensionalization of the former equations results in \cite{murray89}

\begin{eqnarray}
\frac{du}{d\tau}& = & u \left( 1 - u - a_{uv} v \right), \nonumber \\
\\
\frac{dv}{d\tau} &= & \rho v \left( 1 - v - a_{vu} u \right), \nonumber 
\end{eqnarray}

\noindent with the following four equilibria

\begin{eqnarray}
(u^*, v^*) = (0,0),& & (u^*, v^*) = (1,0), \nonumber \\
\label{eq:equil_comp}\\
(u^*, v^*) = (0,1),& &  (u^*, v^*) = \left( \frac{1- a_{uv}}{1-a_{uv}a_{vu}}, \frac{1-a_{vu}}{1-a_{uv}a_{vu}} \right), \nonumber
\end{eqnarray}

\noindent which stability depends solely on the values of $a_{uv}$ and $a_{vu}$, as follows

\begin{itemize}
\item if $a_{uv} < 1$ and $a_{vu} < 1$, there is coexistence, 
\item if $a_{uv} < 1$ and $a_{vu} > 1$, only $u$ survives,
\item if $a_{uv} > 1$ and $a_{vu} < 1$, only $v$ survives,
\item if $a_{uv} > 1$ and $a_{vu} > 1$, either $u$ or $v$ can survive, depending on the initial conditions, but coexistence is forbidden.
\end{itemize}

Diffusion can be also considered, resulting in the following system  

\begin{eqnarray}
\frac{ \partial u}{\partial \tau} &= &  u \left( 1 - u - a_{uv}  v\right) + D_u \nabla^2 u, \nonumber \\
\\
\frac{\partial v}{\partial \tau} &= & \rho v \left( 1 - v - a_{vu} u \right) + D_v \nabla^2 v, \nonumber
\end{eqnarray}
where $D_u$ and $D_v$ are the diffusion coefficients.

As our goal is to include nonlocal interaction terms we propose the following  equations 

\begin{eqnarray}
\frac{ \partial u}{\partial \tau}& = & D_u \nabla^2 u + u \left( 1 - \int_\Omega u(y,t)K_u(|x-y|)dy - a_{uv} \int_\Omega v(y,t)K_v(|x-y|)dy \right),\nonumber  \\
\\
\frac{ \partial v}{\partial \tau}& = & D_v \nabla^2 v + \rho v \left(1 - \int_\Omega v(y,t)K_v(|x-y|)dy - a_{vu} \int_\Omega u(y,t)K_u(|x-y|)dy \right),\nonumber 
\end{eqnarray} 

As mentioned before, the inclusion of the non-local term is related to the fact that at a given time, the  intraspecific and interspecific competition can also be affected by the presence of relatively closed individuals. While it is true that diffusion can help to take into account some of these effects, the interaction between distant individuals will be mediated by the diffusion velocity, while the non-local term can take into account the instantaneous effect that a distant individual may have due to, for example,  a previous presence in the very same place.
In the limiting case when $\omega_i \rightarrow 0$ we recover the local case. 

In the local case, the situation $\rho=1$ with $a_{uv} = a_{vu}$ is not relevant because it actually corresponds to  a single species. However, in the model proposed here, the fact that the non-local range may be different introduces a difference between the otherwise same species. 

Here, we consider a bounded one-dimensional spatial domain of length $2L$, so $\Omega = [-L, L]$. In addition, we consider that $u$ and $v$ are periodic functions in the spatial domain, i.e., with periodic boundary conditions:  $u(-L) = u(L)$ and $v(-L) = v(L)$. To be consistent with the boundary conditions, the kernel function is also $2L$-periodic.  

Considering that we are going to analyze the formation of spatio-temporal patterns in this system, it is worth noticing that even in the case when $D_u\neq D_v$ the conditions to get Turing patterns in the local case are not fulfilled. In fact, a necessary condition is that $1-2u-a_{uv} v$ and $ v ( 1 - v - a_{vu} u)$ had different sign when evaluated at the equilibrium point \cite{murray89}, that does not occur.

\subsection{Results}

We solved the system of equations numerically, considering different combinations of parameters $a_{uv}$, $a_{vu}$, $\omega_u$, $\omega_v$,  and setting, for simplicity, $D_u = D_v=1$ and $\rho=1$. These choices were general enough to unveil all the possible behaviors of the system. To determine the convergence of the numerical method, we calculated the relative difference $\delta_r$ between the curves at two consecutive time steps. 
\begin{equation}
    \delta_r^z=\int_{\Omega} \frac{|z(x,t)-a(x,t-\Delta t)|}{z(x,t-\Delta t)} dx
    \label{dist}
\end{equation}
where $z=\{u,v\}$.
The numerical method was iterated until $\delta_r$  was lower than $10^{-18}$ for each curve.

For the values of $a_{uv}$ and $a_{vu}$, we considered different combinations chosen to scan all the possible scenarios of the long-term behavior of the local system.  For each combination of parameter values, we carried out extensive simulations with  $\omega_u$ and $\omega_v$ in a spatial domain $\Omega = [-500, 500]$ and varying the values of both $\omega$  in steps of 5. As the  spatial domain was discretized in our simulations, we only consider integer values of $\omega_u$ and $\omega_v$.

In agreement with \cite{segal13}, we observe that pattern formation depends on the initial conditions.  Regardless of the formation of patterns,  we find that for some sets of parameter values, the asymptotic behavior of each species also depends on the initial conditions. For example, given a set of parameter values, different initial conditions can or can not lead to a coexistence steady state.
Considering these facts we chose  different initial conditions listed in   Eq. (\ref{eq:IC1}).  As can be seen, the initial conditions are symmetric in the spatial domain.

\begin{eqnarray}
\text{I.C. 1:} & u(x,0) = 0.5 + 0.5 \cdot \cos(2\pi x/100) &\quad v(x,0) = 0.5 + 0.5 \cdot \cos(2\pi x/100) \nonumber \\ 
\text{I.C. 2:} & u(x,0) = 0.5 + 0.5 \cdot cos(2\pi x/100) &\quad  v(x,0) = 0.5 + 0.5\cdot \cos(2\pi x/200)  \label{eq:IC1}  \\
\text{I.C. 3:} & u(x,0) = 0.5 + 0.5 \cdot \cos(2\pi x/100) &\quad  v(x,0) = 0.5 + 0.5\cdot \sin(2\pi (x-25)/100)  \nonumber
\end{eqnarray}

As a general and relevant result it is essential to mention that, given a set of parameters and initial conditions, there is a region in the ($\omega_u,\omega_v$) plane in which the system converges to the homogeneous state without pattern formation. The particular shape and location of this region depend on the values of the other parameters and the initial conditions.  

\subsubsection{Case 1: $a_{uv}<1$ and $a_{vu}<1$} \label{sec:comp_case_1}

\begin{figure}[h]
     \centering 
     \begin{subfigure}[b]{0.495\textwidth}
         \centering
         \includegraphics[width=\textwidth]{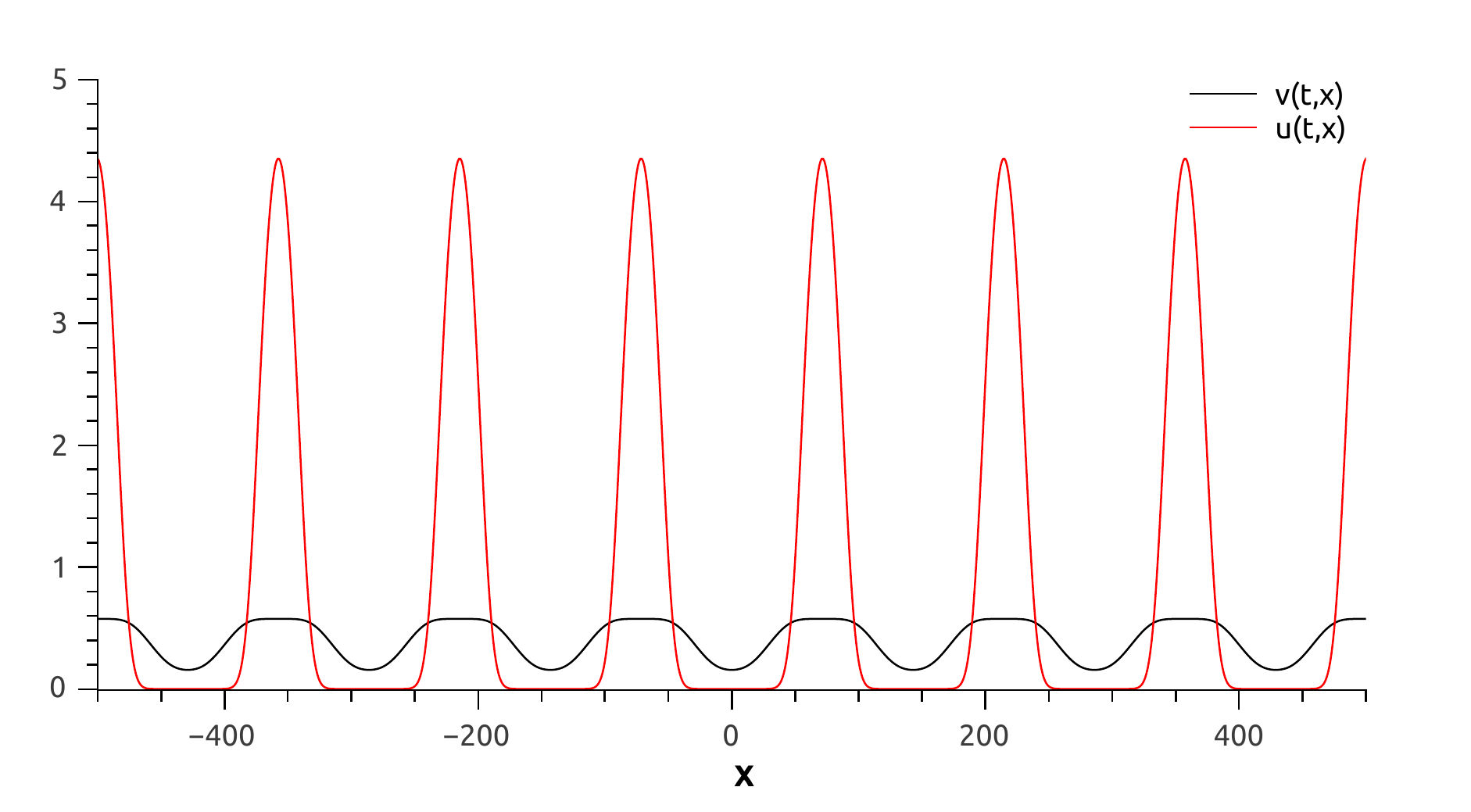}
         \caption{}
         \label{fig:comp_case1_figa}
     \end{subfigure}
     \hfill
     \begin{subfigure}[b]{0.495\textwidth}
         \centering
         \includegraphics[width=\textwidth]{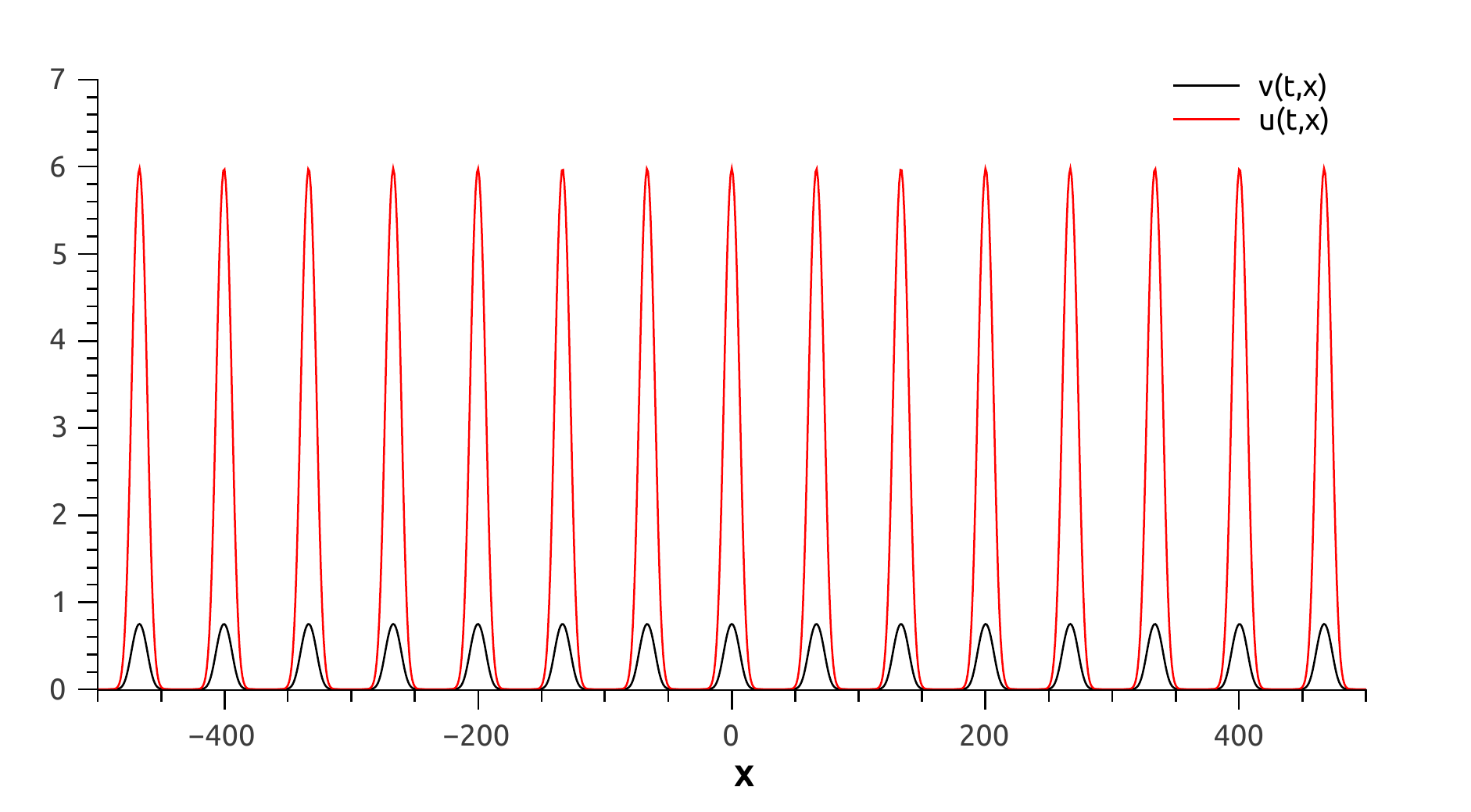}
         \caption{}
         \label{fig:comp_case1_figb}
     \end{subfigure}
    \hfill
     \begin{subfigure}[b]{0.495\textwidth}
         \centering
         \includegraphics[width=\textwidth]{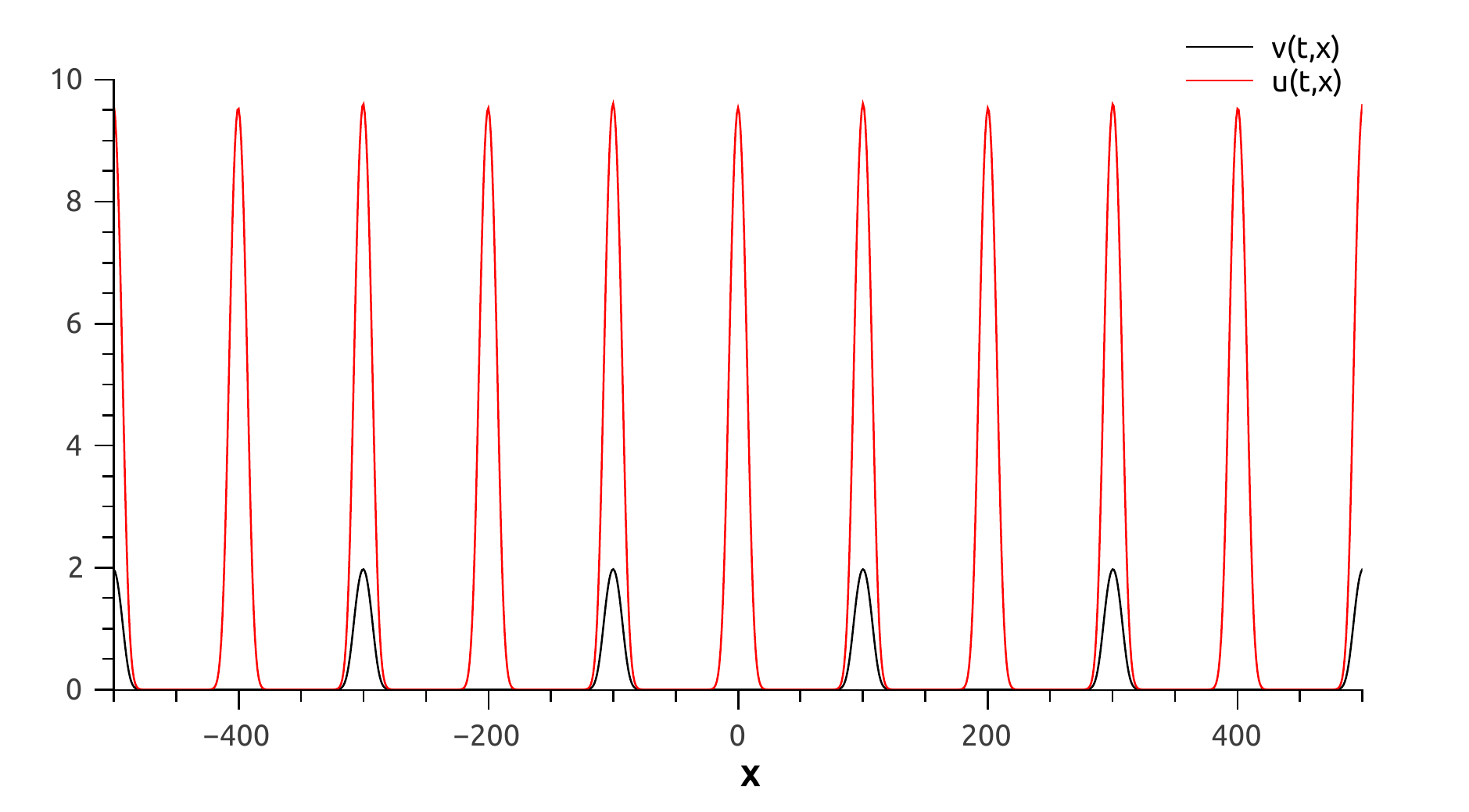}
         \caption{}
         \label{fig:comp_case1_figc}
     \end{subfigure}
      \hfill
     \begin{subfigure}[b]{0.495\textwidth}
         \centering
         \includegraphics[width=\textwidth]{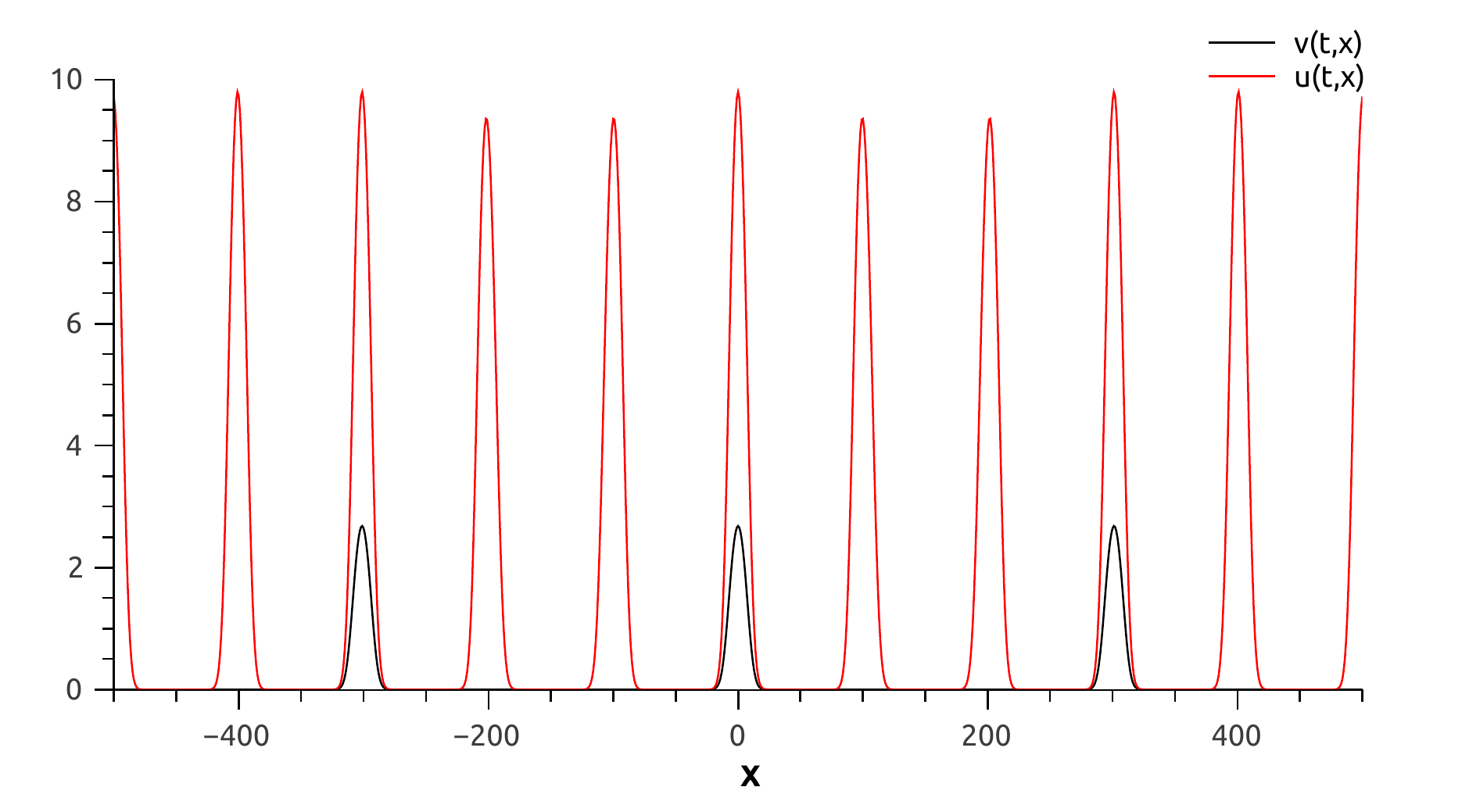}
         \caption{}
         \label{fig:comp_case1_figd}
     \end{subfigure}
     \caption{Spatial pattern formation for: (a) $a_{uv} = 0.5$, $a_{vu} = 0.6$, $\omega_u = 100$, $\omega_v = 0$ and initial condition I.C.3; (b) $a_{uv} = 0.6$, $a_{vu} = 0.9$, $\omega_u = 50$, $\omega_v = 25$ and initial condition I.C.2; (c) $a_{uv} = 0.1$, $a_{vu} = 0.9$, $\omega_u = 80$, $\omega_v = 150$ and initial condition I.C.1; (d) $a_{uv} = 0.1$, $a_{vu} = 0.9$, $\omega_u = 80$, $\omega_v = 200$ and initial condition I.C.1. }
    \label{fig:comp_case1}
\end{figure}

We start with the case corresponding to the occurrence of coexistence in the classical local model. Some examples of the obtained results are shown in Fig (\ref{fig:comp_case1}). We observe that in the non-local model, non-locality does not need to be present in both populations to find spatial patterns in both. Indeed, pattern formation also occurs when non-locality is present only in one of the species (Fig. \ref{fig:comp_case1_figa}). Furthermore, when both species present non-locality, the spatially patterned region of survival can be overlapped in the entire domain with their peaks at the same $x$ values (Fig. \ref{fig:comp_case1_figb}), or partially overlapped with different periodicity as in (Fig. \ref{fig:comp_case1_figc}) and (Fig. \ref{fig:comp_case1_figd}).

While the values of $\omega_v$ and $\omega_u$ do not determine the maximum value $u_{\text{max}}$ and $v_{\text{max}}$ we can observe that when  $D_v=D_u$, $a_{uv}=a_{vu}$ and $\rho=1$, we will have $u_{\text{max}} \geq v_{\text{max}}$ if $\omega_u > \omega_v$.  In addition, if $a_{uv} = a_{vu}$ and $\omega_u = \omega_v$, the global distribution of the species in space may be identical even though the initial conditions are different. On the other hand, if $a_{uv} \neq  a_{vu}$, the global distribution of the species is not identical in any case.

Although in all the cases presented here (with $a_{uv} < 1$ and $a_{vu} <1)$ we can observe the presence of both species in space, we can find regions of coexistence sometimes intercalated with empty regions or regions where only  one of the species survives (always the same) as in Figs. (\ref{fig:comp_case1_figc}) and (\ref{fig:comp_case1_figd}). However, we do not attribute this feature to some sort of local competitive exclusion but to the patterns associated with the different values of $\omega_i$. In summary,  when $a_{uv} < 1$ and $a_{vu} < 1$, we did not found situations of competitive exclusion. 

In addition, for a given set of parameters and initial conditions, there is a region of $\omega_u$ and $\omega_v$ ($R_h$) in which, the non-local model converges to a homogeneous equilibrium with the values given by Eq. \ref{eq:equil_comp}. For example, let us consider $a_{uv} = 0.1$ and $a_{vu}=0.9$ and call this region  $R_{h,1} = \{(\omega_u, \omega_v): 0 \leq \omega_u \leq 5,  0 \leq \omega_v \leq 25 \}$. Inside this region, the model reaches a homogeneous steady state (or with patterns with amplitude in the order of $10^{-14}$), for the three initial conditions analyzed here. This region is not exhaustive and may exist other combinations of $\omega_u$ and $\omega_v$ that produce homogeneous steady states outside this region. 

\subsubsection{Case 2: $a_{uv}>1$ and $a_{vu}>1$} \label{sec:comp_case_2}

\begin{figure}[h]
     \centering 
     \begin{subfigure}[b]{0.495\textwidth}
         \centering
         \includegraphics[width=\textwidth]{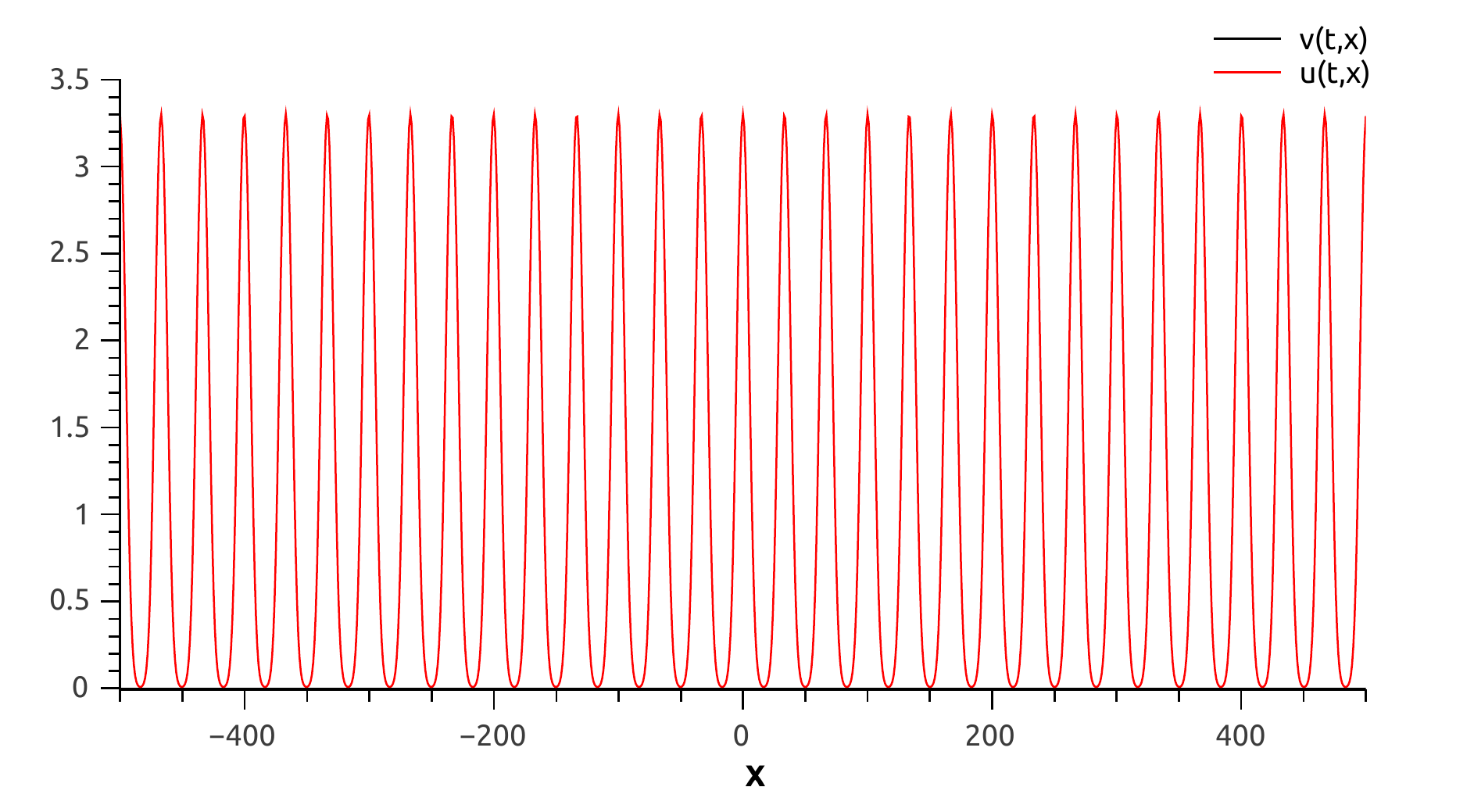}
         \caption{}
         \label{fig:comp_case2_figa}
     \end{subfigure}
     \hfill
     \begin{subfigure}[b]{0.495\textwidth}
         \centering
         \includegraphics[width=\textwidth]{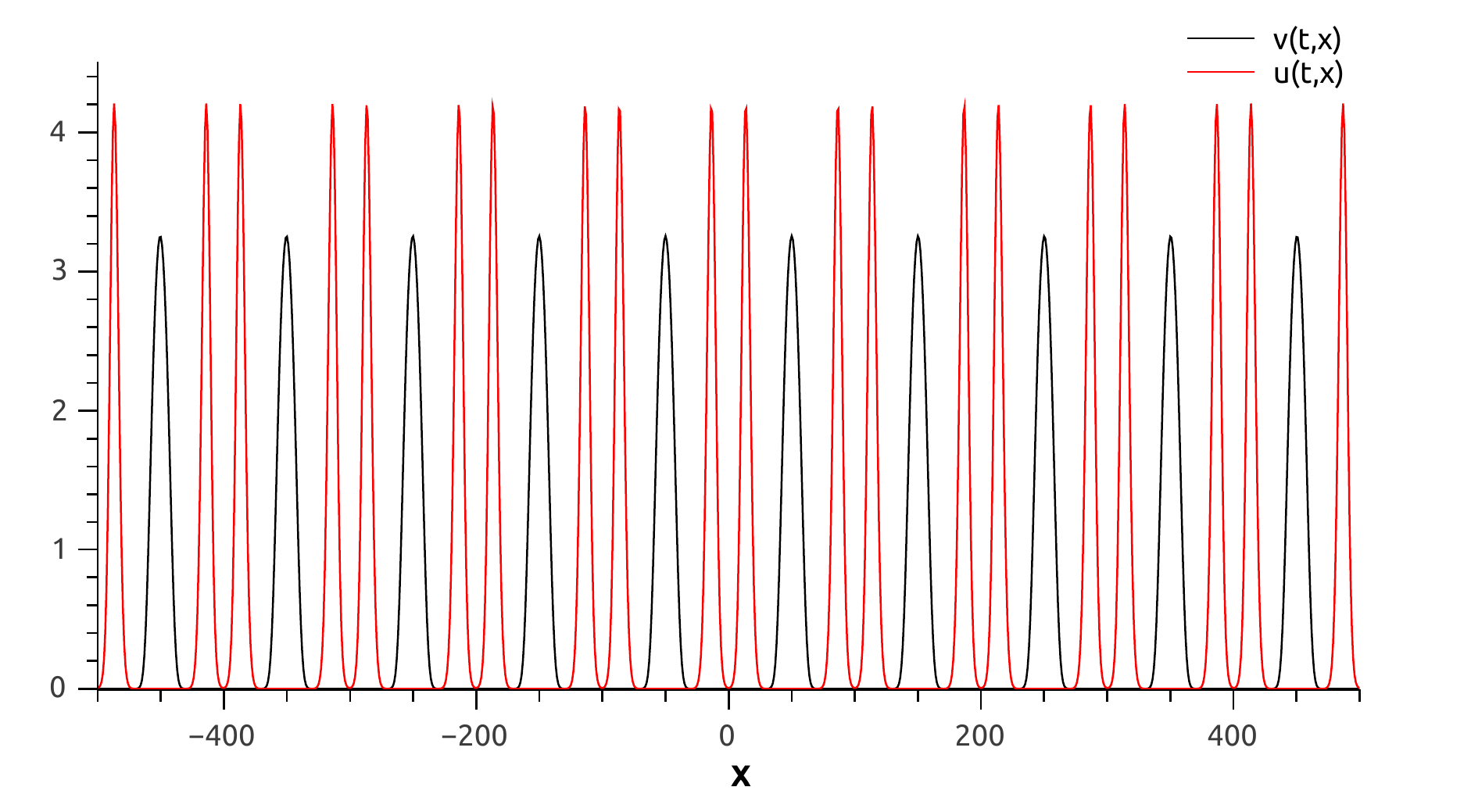}
         \caption{}
         \label{fig:comp_case2_figb}
     \end{subfigure}
    \hfill
     \begin{subfigure}[b]{0.495\textwidth}
         \centering
         \includegraphics[width=\textwidth]{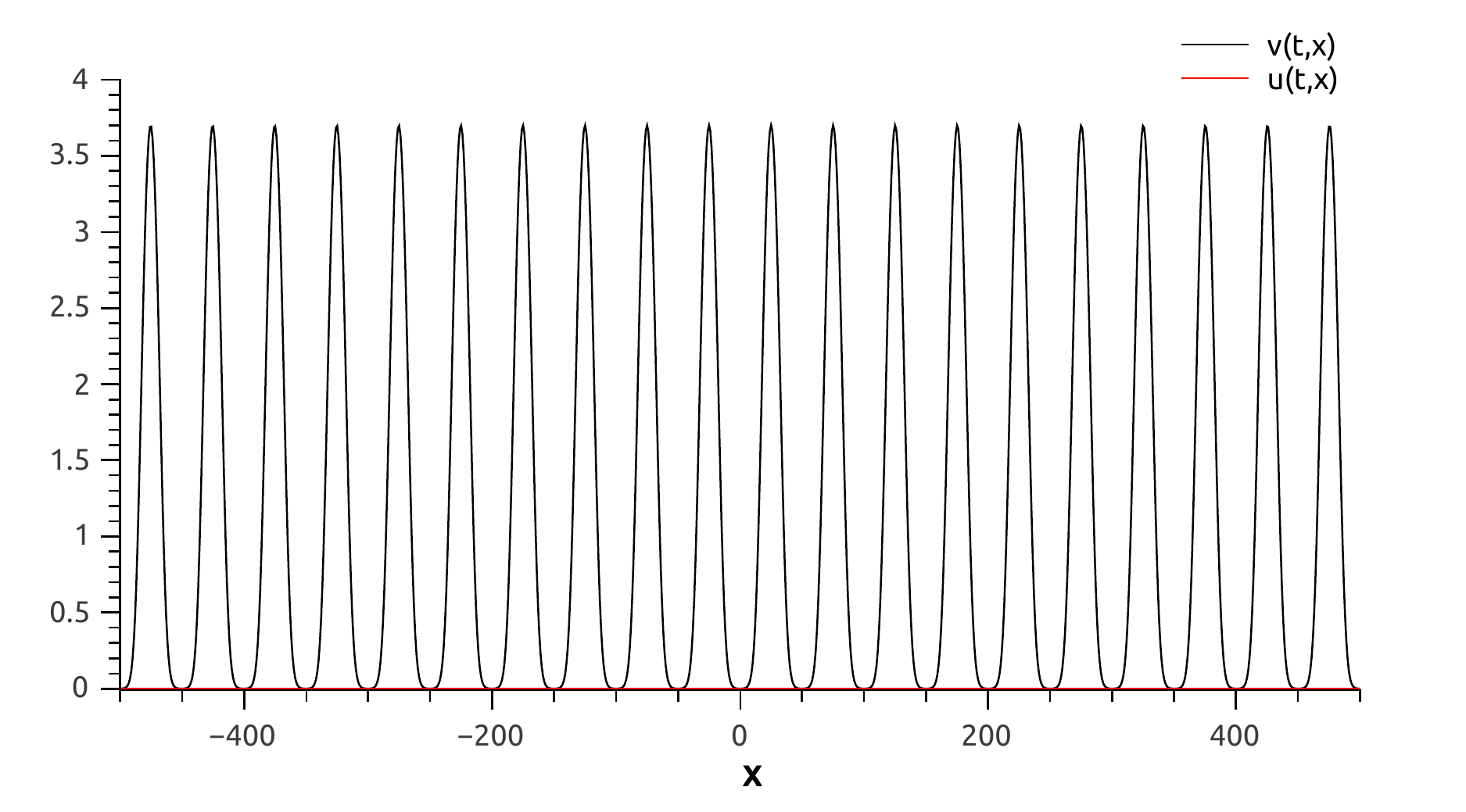}
         \caption{}
         \label{fig:comp_case2_figc}
     \end{subfigure}
     \caption{Spatial pattern formation for $a_{uv}=1.5$, $a_{vu}=2$, initial condition I.C.3, $\omega_u = 20$ and (a) $\omega_v = 15$; (b) $\omega_v = 25$; (c) $\omega_v = 30$.}
    \label{fig:comp_case2}
\end{figure}

 In the local model, when $a_{uv}>1$ and $a_{vu}>1$ there is no coexistence of species, and what determines which species survives are the initial conditions. There are examples in which the inclusion of diffusion allows the coexistence of the species \cite{wio1,wio2} but certain special conditions must be met. Here we show that if non-local interactions are considered, it is possible to achieve competitive coexistence when otherwise it would be prohibited. Some of the results are shown in Fig. (\ref{fig:comp_case2}).

For example, if we consider $a_{uv}=a_{vu}$ and the same initial conditions (I.C.1), we observe that for some values of the ranges $\omega_u $ and $\omega_v$, the scenario can be interesting, presenting islands of local competitive exclusion such that the regions colonized by each species are intertwined. Global exclusion is observed for other values of $\omega_u$ and $\omega_v$. It is interesting to note that a larger $\omega$ value does not necessarily result in a competitive advantage over the other species. For example, considering $a_{uv} = a_{vu} = 1.1$, $\omega_u = 25$, $\omega_v = 35$ and initial condition I.C.1, the species $u$ survives and $v$ becomes extinct. The same occurs for $\omega_v = 40$, and $\omega_v = 45$, for example.

If the initial distributions of the species are different, it would not seem that an identical coexistence occurs, and the cases of coexistence are of a global nature, but associated with local competitive exclusion even when $\omega_u = \omega_v$. That is, both species are present but occupy different regions and are never both in the same place.

In another concrete example, when $a_{uv} \neq a_{vu}$ ($a_{uv} = 1.5$ and $a_{vu}=2$), and the initial conditions are the same for both species (I.C.1), there is no coexistence and only the species $u$ survives, regardless of the values of $\omega_u$ and $\omega_v$. If the initial conditions are different, for the considered parameters ($a_{uv} = 1.5$, $a_{vu}=2$, $D_v=1$, and $\rho=1$), coexistence is not commonly observed, and can be found for only a few combinations of $\omega_u$ and $\omega_v$. In this situation, as $\omega_u$ and $\omega_v$ change, we observe an alternation of  global exclusion and coexistence. An interesting aspect is that each coexistence region is adjacent to two exclusion regions such that in each of them the surviving species is different. However, these two exclusion regions can be adjacent without a coexistence region in between. An example of the coexistence region in the middle, is the following: Consider the initial condition I.C.3 and $\omega_u = 20$, if $\omega_v = 15$, the species $u$ survives (Fig. \ref{fig:comp_case2_figa}); if $\omega_v = 20$ there is coexistence, the same as if $\omega_v = 25$ (Fig. \ref{fig:comp_case2_figb}); and if $\omega_v = 30$, the surviving species is $v$ (Fig. \ref{fig:comp_case2_figc}). As before, all cases of global coexistence are associated with local competitive exclusion.

As before, we can find a region for $\omega_u$ and $\omega_v$ for which the system converges to a steady state without spatial patterns. In this case, this region is very sensitive to the initial conditions. If we considered $a_{uv} = 1.5$ and $a_{vu}=2$ and the same initial conditions for both species (I.C.1), we found that inside the region $R_{h} = \{(\omega_u, \omega_v): 0 \leq \omega_u \leq 5, 0 \leq \omega_v \}$ the system converges to a homogeneous state where only the specie $u$ survives. However, when the initial conditions are different (I.C.2 and I.C.3.) we had to reduce that region to $R_{h} = \{(\omega_u, \omega_v): 0 \leq \omega_u \leq 5, 0 \leq \omega_v \leq 5 \}$ to not have pattern formation.

\subsubsection{Case 3: $a_{uv} > (<)1$ and $a_{vu}<(>)1$} \label{sec:sp-temp-competition}

\begin{figure}[h]
    \centering
    \includegraphics[width=0.75\textwidth]{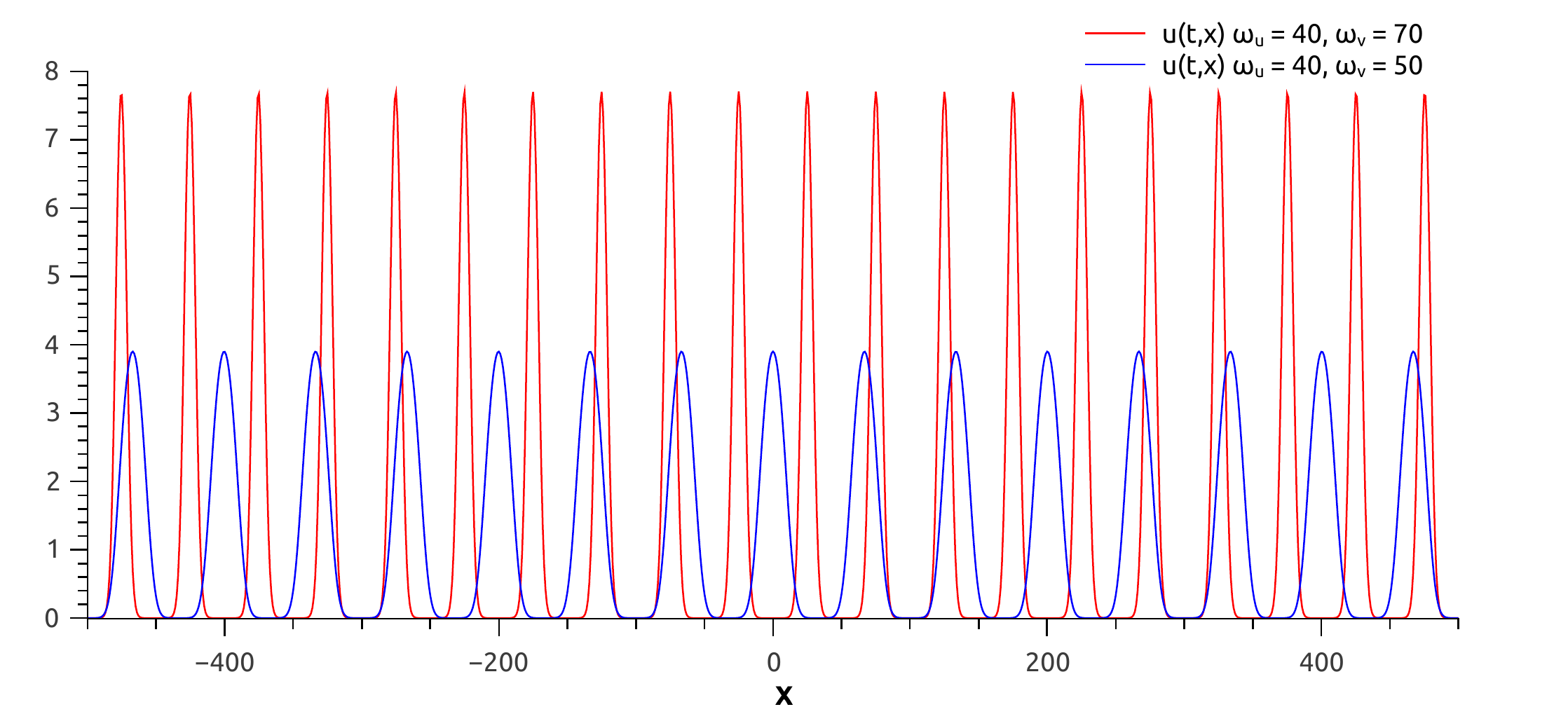}
    \caption{Spatial pattern formation for $a_{uv}=0.9$, $a_{vu}=1.1$, initial condition I.C.2, $\omega_u = 40$ and: $\omega_v = 70$ (blue) and $\omega_v = 50$ (red).}
    \label{fig:comp_case3}
\end{figure}

In this case, the inclusion of the non-locality in the model (with the parameters and initial conditions that we considered) does not generate results different from those obtained in the local case in terms of coexistence or exclusion. That is, only one species survives. The only new feature is that the system converges to a non-homogeneous state for the survivor, which corresponds to the species with $a_{i,j} <1$. It is interesting to mention what we called a {\it phantom effect} associated with the fact that  despite the extinction of one of the species, its previous presence, and its corresponding $w$ range affect and determine the final distribution of the surviving species (Fig. \ref{fig:comp_case3}).  If we considered $a_{uv} = 0.9$ and $a_{vu} = 1.1$, a region for which the system reaches a homogeneous state is $R_{h} = \{(\omega_u, \omega_v): 0 \leq \omega_u \leq 5, 0 \leq \omega_v \}$ for the three initial conditions analyzed, in a non-exhaustive way.

\subsection{Spatio-temporal patterns}
One of the rationales to introduce the distance $\delta^r$, Eq. (8), was to test the convergence of the system to a steady state. But also we wanted to check the possibility of temporal patterns that would be reflected in oscillations of $\delta^r$. 

During the extensive simulations carried out for the parameters $\omega_u$ and $\omega_v$, we noticed a range of parameters for which the differences between curves have indeed a periodic behavior.  While in the local model it is not possible to obtain periodic solutions, we verified their existence in the non-local model, always associated with spatio-temporal patterns. 

To deepen in the characterization of this phenomenon, we considered the region in the parameter space determined by $\omega_u = 10$, $10 \leq \omega_v \leq 20$, $1.05 \leq a_{uv} \leq 1.15$, $1.05 \leq a_{vu} \leq 1.15$, with the initial condition 2 (Eq. \ref{eq:IC1}). This region was analyzed considering a step of $0.01$ for parameters $a_{uv}$ and $a_{vu}$, and a step of $1$ for $\omega_v$. As a result, we obtained different combinations of parameters for which a periodic behavior is observed. These combinations are summarized in Fig. \ref{fig:STP-parameters}, and a detail for each $\omega_v$-level can be seen in the Supplementary Material (Sec. \ref{sec:sm-stpattern}). In this case, we found that if $\omega_v < 13$ there are no periodic solutions in time, and that $\omega_v = 15$ is the level that has more periodic solutions in  time. 

\begin{figure}[h]
    \centering
    \includegraphics[width=0.65\textwidth]{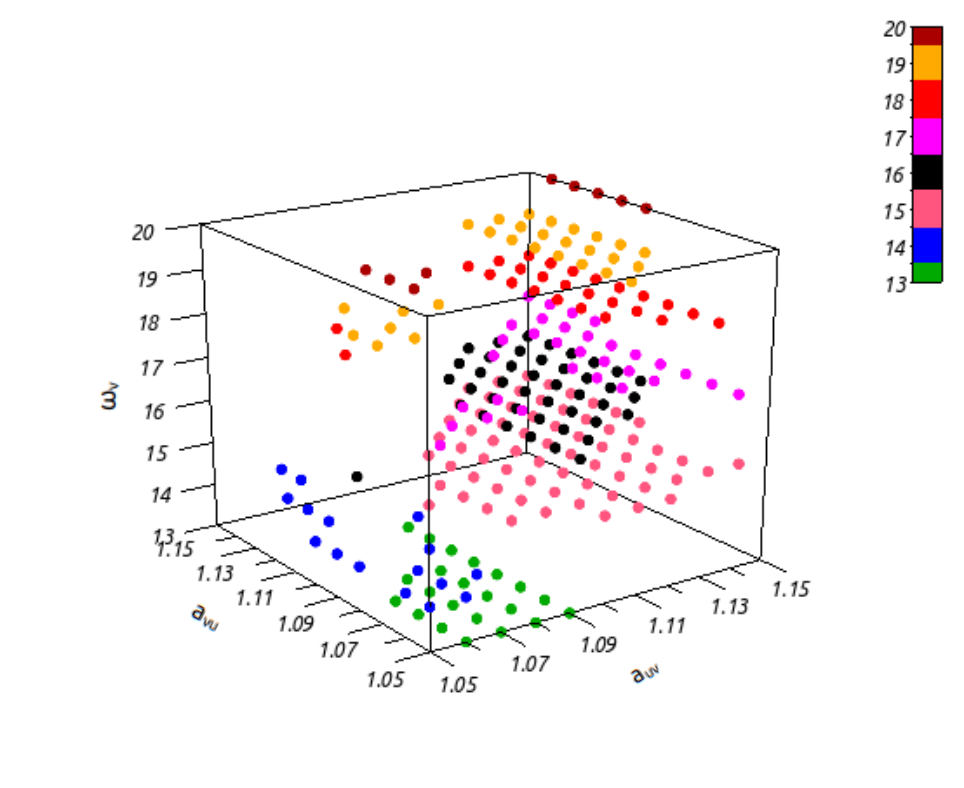}
    \caption{Combinations of $a_{uv}$, $a_{vu}$ and $\omega_v$ for which spatio-temporal pattern is observed, considering $\omega_u = 10$ and initial condition 2 (Eq. \ref{eq:IC1}).}
    \label{fig:STP-parameters}
\end{figure}

\begin{figure}[h]
     \centering 
     \begin{subfigure}[b]{0.495\textwidth}
         \centering
         \includegraphics[width=\textwidth]{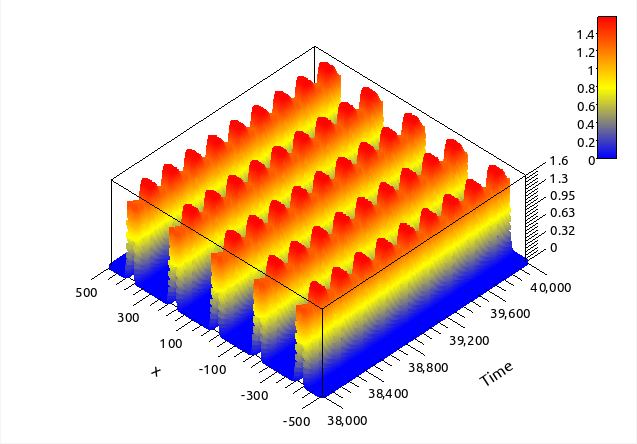}
         \caption{}
     \end{subfigure}
     \hfill
     \begin{subfigure}[b]{0.495\textwidth}
         \centering
         \includegraphics[width=\textwidth]{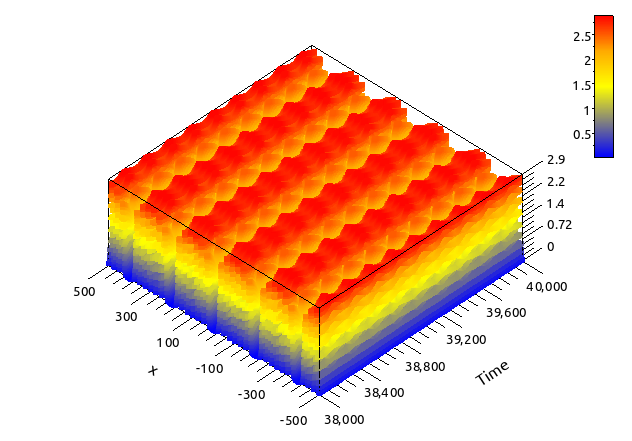}
         \caption{}
     \end{subfigure}
    \hfill
     \begin{subfigure}[b]{0.495\textwidth}
         \centering
         \includegraphics[width=\textwidth]{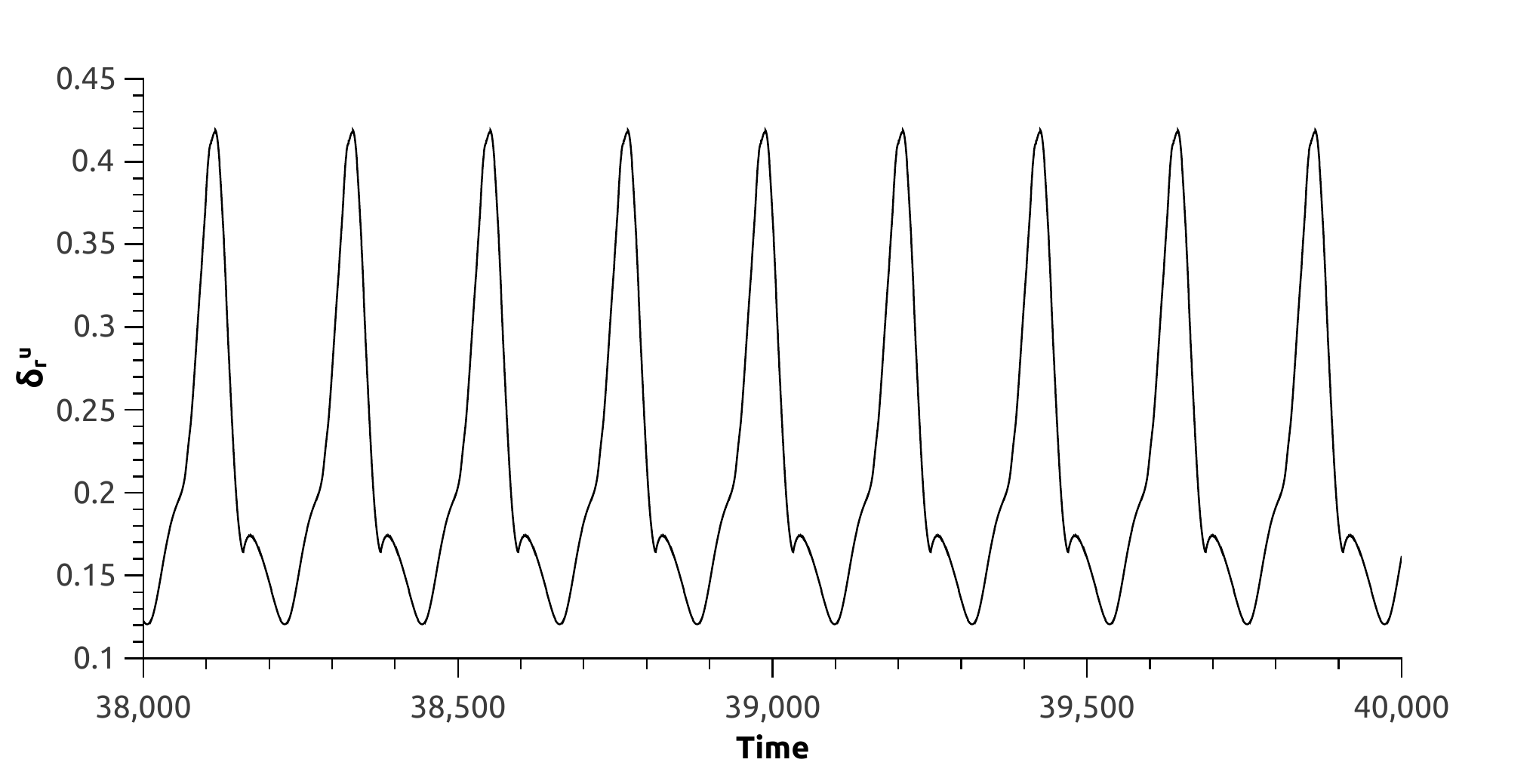}
         \caption{}
     \end{subfigure}
      \hfill
     \begin{subfigure}[b]{0.495\textwidth}
         \centering
         \includegraphics[width=\textwidth]{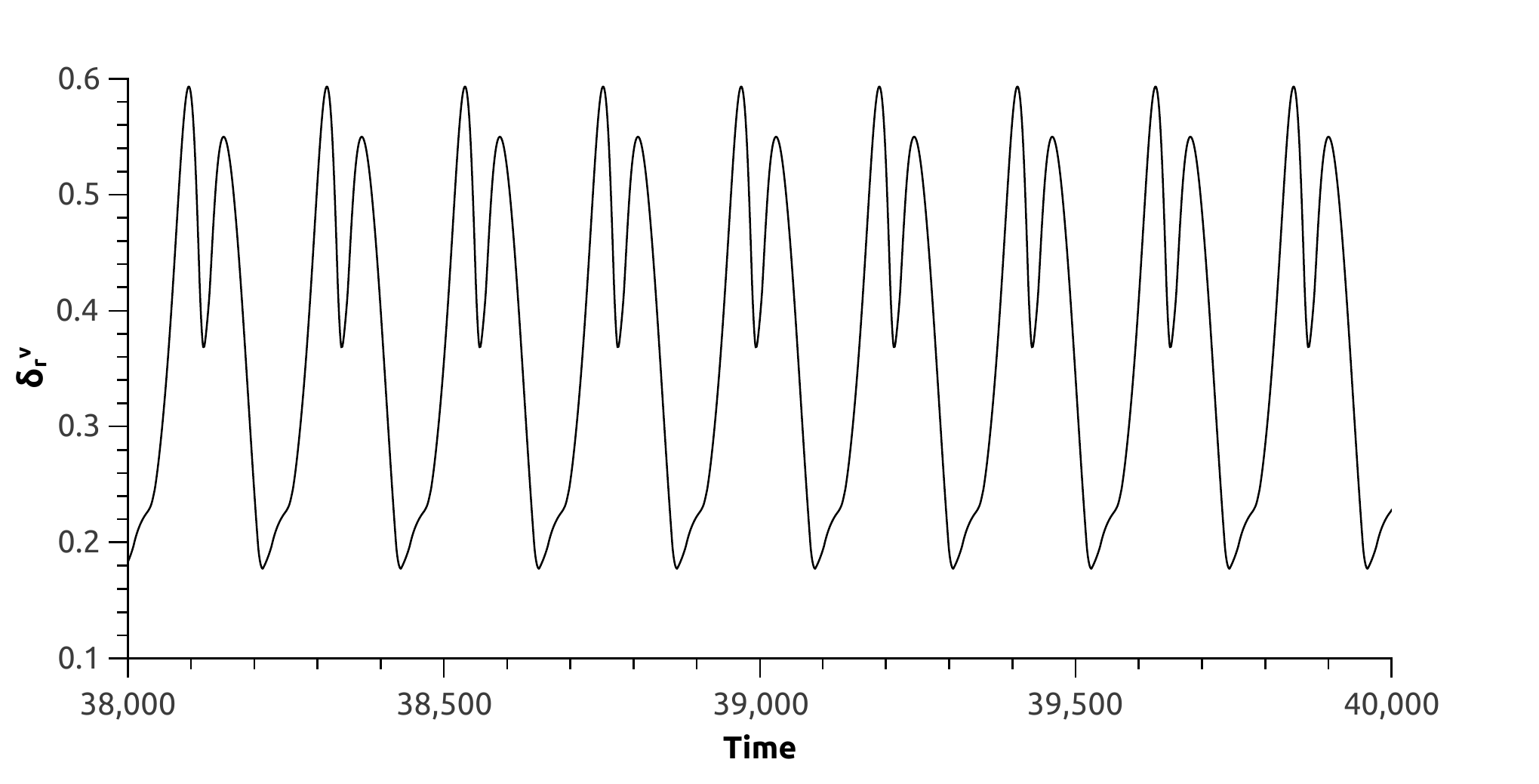}
         \caption{}
     \end{subfigure}
         \caption{Spatio-temporal patterns obtained for (a) species $u$  and (b) $v$ considering $\omega_u = 10$, $omega_v=15$, $a_{uv}=1.12$, $a_{vu}=1.08$ and the initial condition 2. In (c) and (d) the differences according to Eq. \ref{dist} for $u$ and $v$, respectively.}
    \label{fig:STP-pattern}
\end{figure}

To illustrate this situation, we show in Fig. \ref{fig:STP-pattern} the spatio-temporal patterns obtained considering: $\omega_u = 10$, $\omega_v=15$, $a_{uv}=1.12$, $a_{vu}=1.08$ and the initial condition 2.

 We mentioned that in some cases, we observed local competitive exclusion but global coexistence. When also temporal patterns appear the species try to preserve their own space, and the oscillations are greater in the boundaries of the regions that delimited the areas colonized by each species. There, at this interface, the competition is effective and the oscillations are larger than in the rest of the domain. For example, in Fig. \ref{fig:STP-pattern} we can see that the species are grouped into local isolated islands over time, and the oscillations appear at the boundaries. If the islands are thin, as is the case with species $u$, the entire island is observed to oscillate.
 Spatio-temporal patterns were not detected in cases where only one specie survives.

While the results presented in this section only occur within a very restricted region of parameters, the results were tested for consistency and the possibility of numerical artifacts was ruled out. For example, in the case shown above, when we considered $a_{uv}=1.12$, $a_{vu}=1.08$ and the initial condition 2, the percentage of $\omega_u$-$\omega_v$ combinations that could produce spatio-temporal patterns is around 1\%.

\section{Predator-prey interaction}

The second case to be analyzed corresponds to  a two-species system with one prey, $p$, and one predator, $q$. We propose logistic demography and a predation term with saturation for the prey, as in \cite{murray89}. Further, we assume that the dynamics of the predator presents a carrying capacity proportional to the prey population, $K_q\propto p$. The equations, already adimensionalized, read:   

\begin{eqnarray}
\dot{p} &=& p \left(1-p\right) -a \frac{p q}{p+b}, \nonumber \\
\\
\dot{q} &= & c\,q  \left(1-\frac{q}{p}\right). \nonumber 
\end{eqnarray}
\begin{eqnarray}
&& \nonumber
\end{eqnarray}

This system has only one relevant equilibrium 
\begin{equation}
p^*=q^*=\frac{1-a-b+\sqrt{(1-a-b)^2+4b}}{2}
\label{eq:equil}
\end{equation}
and presents two different behaviors clearly associated with  two regions in the parameter space.
If the condition 
\begin{equation}
c>[a-\{(1-a-b)^2+4b\}^{1/2}]\frac{[1+a-b-\{(1-a-b)^2+4b\}^{1/2}]}{2a}
\label{condT1}
\end{equation}
is fulfilled then the equilibrium is stable, otherwise, the system oscillates due to the presence of  a limit cycle.

After the inclusion of diffusion and non-local effects in the interaction terms, we can write

\begin{eqnarray}
\frac{ \partial p}{\partial t}& = & D_p \nabla^2 p + p \left(1-\int_\Omega p(y,t)K_p(|x-y|)dy \right) -a \frac{p\int_\Omega q(y,t)K_q(|x-y|)dy}{\int_\Omega p(y,t)K_p(|x-y|)dy+b}, \nonumber \\
& & \\
\frac{ \partial q}{\partial t}& = & D_q \nabla^2 q+ c\,q  \left(1-\frac{\int_\Omega q(y,t)K_q(|x-y|)dy }{\int_\Omega p(y,t)K_p(|x-y|)dy}\right), \nonumber
\end{eqnarray}

\noindent where we have considered that no locality affects the interaction terms and that the carrying capacity of the predator is proportional to the amount of prey within a range around the location of each individual.  As before, we consider a one-dimensional spatial domain of length $2L$, periodic boundary conditions for $p$ and $q$, and the $2L$ periodic kernel function defined in Eq. \ref{eq:kernel}.

We can check again whether the condition for the possibility of Turing patterns is fulfilled or not.
In the present case, we need 
\begin{equation}
1 - 2 p^* - \frac{a b q^*}{(b + p^*)^2}>0.
\label{condT2}
\end{equation}
Replacing $p^*$ and $q^*$ in Eq. (\ref{condT2}) we get 
\begin{equation}
b < \sqrt{4 a + a^2}-1-a.
\label{condT3}
\end{equation}
In the limiting case, when $b=\sqrt{4 a + a^2}-1-a $, we can verify that the stability condition, Eq. (\ref{condT1}), is equivalent to 
$$c>0,$$ that is always true, meaning that the surface separating the regions where Turing patterns are possible from those where they are not is well within the stability region. That means that we can find a set of parameters for which the equilibrium is stable and Turing patterns are forbidden.

\subsection{Results}

 As in the previous case, the system was solved numerically by setting $D_v = D_u = 1$, and considering different combinations for $a$, $b$, and $c$, in order to analyze the situation inside and outside of the stability region in the local case. For the first case, we considered two parameter combinations: comb1: $a=1$, $b=0.2$, $c=0.8$, and comb2: $a=2$, $b=0.4$, $c=0.1$; while for the second case we use $a=0.8$, $b=0.05$ and $c=0.25$. For the cases within the stability region, according to the parameters considered Turing patterns are not possible. We also considered different initial conditions, 
 \begin{eqnarray}
\text{I.C. 1:} & u(x,0) = 0.6 + 0.5 \cdot \cos(2\pi x/200) &\quad v(x,0) = 0.6 + 0.5 \cdot \cos(2\pi x/200) \nonumber \\ 
\text{I.C. 2:} & u(x,0) = 0.6 + 0.5 \cdot cos(2\pi x/100) &\quad  v(x,0) = 0.6 + 0.5\cdot \cos(2\pi x/200)  \label{eq:IC-pp}  \\
\text{I.C. 3:} & u(x,0) = 0.6 + 0.5 \cdot \cos(2\pi x/200) &\quad  v(x,0) = 0.6 + 0.5\cdot \sin(2\pi (x-50)/200)  \nonumber
\end{eqnarray}
and carried out extensive simulations for the values of $\omega_p$ and $\omega_q$ in a spatial domain $\Omega = [-500,500]$.

\subsubsection{Case 1: within the stability region of the local model.}

In the local model, the system has only one relevant equilibrium given by Eq. \ref{eq:equil}. In the cases analyzed here, we found a region of $\omega_p$ and $\omega_q$ ($R_h$) in which, the non-local model also converges to a homogeneous equilibrium with the values given by Eq. \ref{eq:equil}. This region depends on the values of $a$, $b$, and $c$ and the initial conditions. For the combination of parameters comb1, when $(\omega_p, \omega_q)$ are in the region $R_{h,1} = \{(\omega_p, \omega_q): 0 \leq \omega_p \leq 10,  \omega_q < 0.5 \cdot \omega_p + 5 \}$, the system reaches a homogeneous steady state though some times  we observed  spatial patterns with an amplitude in the order of our numerical resolution. This happens for the three initial conditions analyzed here.

A similar situation occurs when the combination of parameters comb2 is considered. In this case, a region within which the system converges to a homogeneous state (or with amplitudes in the limit of our numerical resolution) can be defined as $R_{h,2} = \{(\omega_p, \omega_q): 0 \leq \omega_p \leq 10,  \omega_q \leq 0.5 \cdot \omega_p + 5 \} \cup \{ (\omega_p, \omega_q): \omega_p = \omega_q, 10 \leq \omega_p \leq 50\}$. 

The regions defined below are not exhaustive, that is, near these regions, other combinations of $\omega_p$ and $\omega_q$ can cause the system to converge to a homogeneous state, however, this occurs for only a few combinations of these parameters.  

However, in most cases, non-locality produces pattern formation, and when this happens we observed two different situations: the system can converge to a temporally stable spatial pattern, or it can present spatio-temporal patterns.

\begin{figure}[h]
     \centering 
     \begin{subfigure}[b]{0.495\textwidth}
         \centering
         \includegraphics[width=\textwidth]{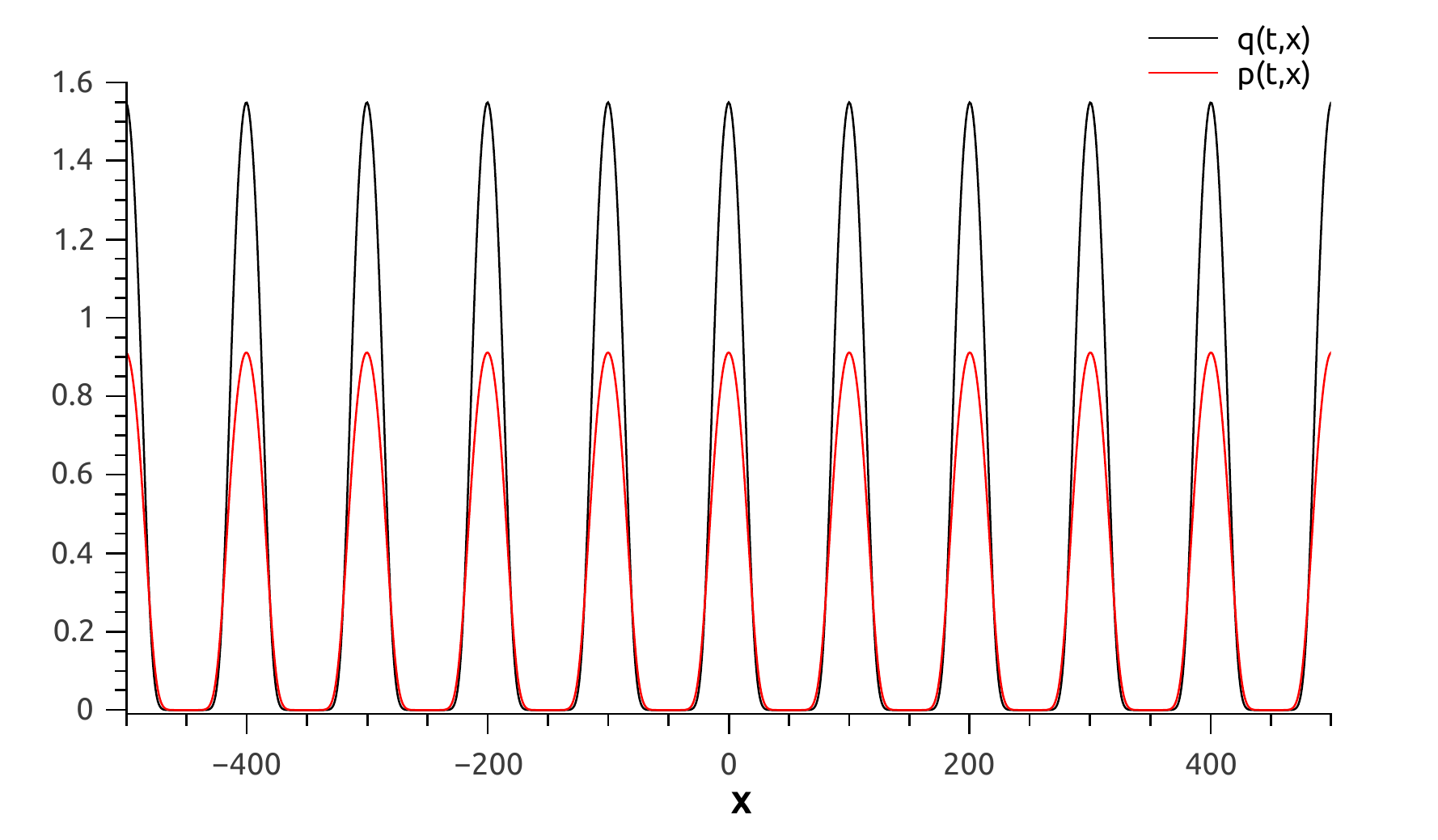}
         \caption{}
         \label{fig:cond_omegas_a}
     \end{subfigure}
    \hfill
     \begin{subfigure}[b]{0.495\textwidth}
         \centering
         \includegraphics[width=\textwidth]{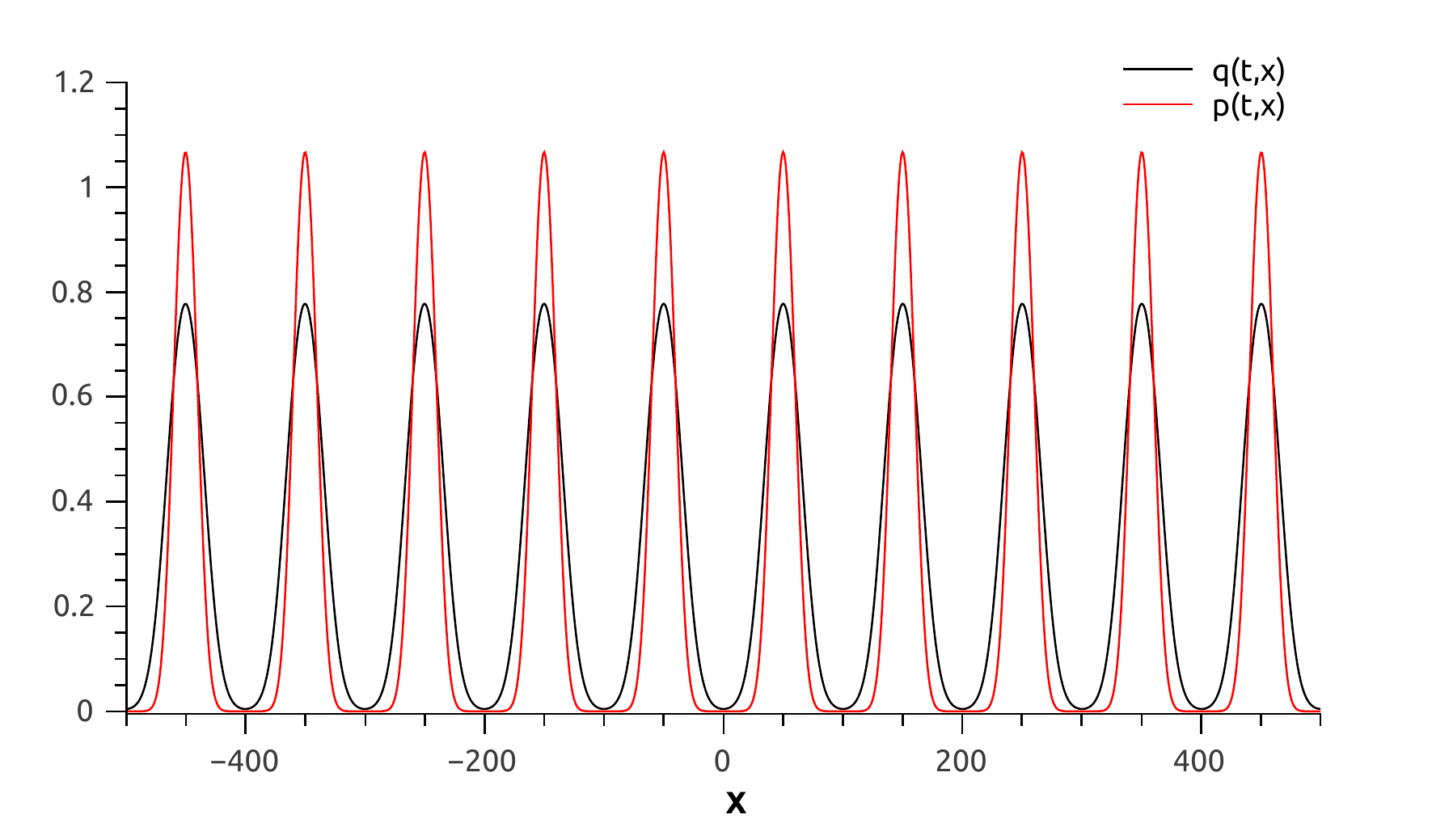}
         \caption{}
         \label{fig:cond_omegas_b}
     \end{subfigure}
         \hfill
     \begin{subfigure}[b]{0.495\textwidth}
         \centering
         \includegraphics[width=\textwidth]{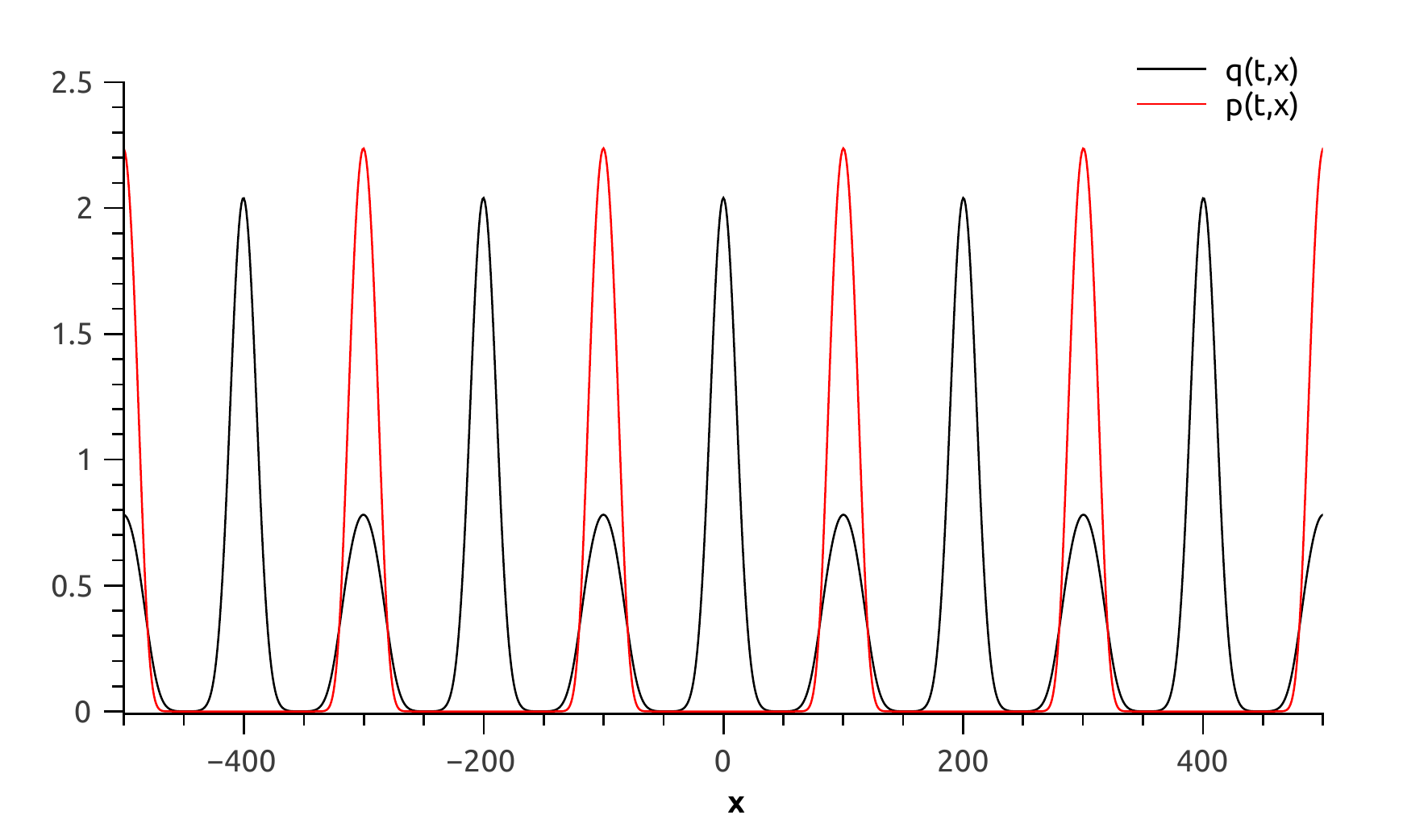}
         \caption{}
         \label{fig:cond_omegas_c}
     \end{subfigure}
     \caption{Temporally stable spatial pattern considering: (a) comb1, $\omega_p=40$, $\omega_q=60$ and initial condition I.C.2; (b) comb2, initial condition I.C. 2, $\omega_p = 50$, $\omega_q = 60$; (c) comb2, initial condition 3, $\omega_p=120$, $\omega_q=60$.}
    \label{fig:cond_omegas}
\end{figure}

When species reach a temporally stable spatial pattern, the spatially patterned region of survival may or may not be overlapped in the entire domain with their peaks at the same $x$ values. For both comb1 and comb2, we find that a sufficient condition for patterns to be overlapped in the entire domain with their peaks at the same $x$ values is that $\omega_p < \omega_q$. Even more, for comb1, when this condition is fulfilled, the temporally stable spatial pattern satisfies that the height of the peaks of $p$ is less than or equal to those of $q$ and the same happens with the abundance of species, that is, $\int_\Omega p(x,t)dx \leq \int_\Omega q(x,t)dx$ (Fig. \ref{fig:cond_omegas_a}). However, while the statement about abundances is generally true when comb2 is considered, with only a few counterexamples (in which the abundance of u exceeds that of v by less than 5\%), we can find many counterexamples to the statement about maxima (Fig. \ref{fig:cond_omegas_b}). If the condition $\omega_p < \omega_q$ is not fulfilled, the spatially patterned region of survival may not be overlapped in the entire domain, and $q$ can colonize region where the $p$ is absent (Fig. \ref{fig:cond_omegas_c}). This last result shows that non-locality allows the predator to survive in regions where the prey is not present, as long as the interaction range of the prey is greater than that of the predator. This situation leads to a paradoxical result. As can be seen in Fig. \ref{fig:cond_omegas_c}, predator abundance peaks in the regions not shared with the prey  are higher than those that are in the coexistence region. The apparent paradox can be explained due to the non-local effects.  The predators that are in the regions not shared with the prey can feed on the two adjacent prey locations, having twice more  available resources than the predators that are in the coexistence regions, which scope is limited to this area.

\subsubsection{Spatio-temporal patterns within the stability region of the local model}

Within the stability region of the local model, it is not possible to find periodic solutions when non-locality is not considered. However, similarly to what happened for the competition model (Sect. \ref{sec:sp-temp-competition}), there are values of $\omega_p$ and $\omega_q$ for which the non-local model presents spatial-temporal patterns. As before, these values depend on combinations of parameters $a$, $b$, and $c$, and the initial conditions. 

For the combinations comb1 and comb2, we observed that these patterns can be found on or near the diagonal $(\omega_p, \omega_q)$, always outside the regions $R_{h,1}$ and $R_{h,2}$, respectively. However, this characterization is not exhaustive, and other $\omega_p$ and $\omega_q$ combinations that generate spatio-temporal patterns can be found. 

\begin{figure}[h]
    \centering
    \includegraphics[width=0.5\textwidth]{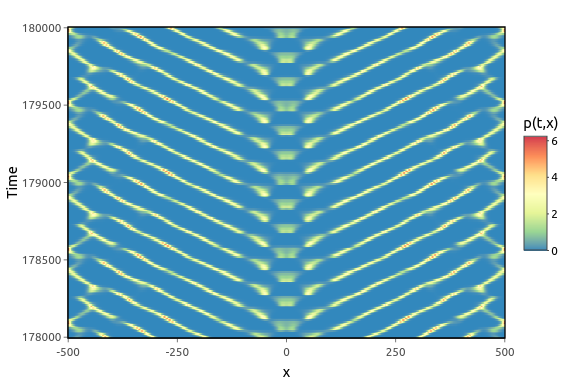}\includegraphics[width=0.5\textwidth]{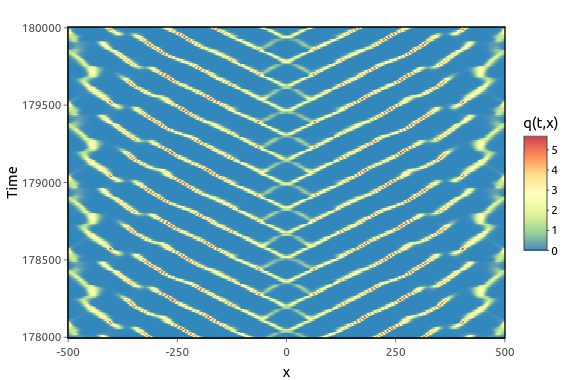}
    \caption{Spatio-temporal patterns obtained for species $p$ (left) and $q$ (right) considering $\omega_p=100$, $\omega_q=95$, $a=1$, $b=0.2$, $c=0.8$ and the initial condition 2.}
    \label{fig:temp_pat_pq}
\end{figure}

To illustrate this situation, consider $a=1$, $b=0.2$, $c=0.8$ (comb1), the initial condition I.C.2, $\omega_p = 100$ and $\omega_q = 95$. The long-term behavior for this case is in Fig. \ref{fig:temp_pat_pq}. We can see clearly the presence of spatio-temporal patterns. 

It is important to mention that inside the stability region of the local model, the most common long-term behavior is the convergence to a temporally stable spatial pattern. For example, when comb1 and initial condition I.C. 2 are considered, less than 5\% of combinations of $\omega_p$ and $\omega_q$ produce spatio-temporal patterns.

\subsubsection{Case 2: outside the stability region of the local model}

As in the previous cases, we found a region in the $\omega_p-\omega_q$ plane within which the system reaches a homogeneous state. For the parameters $a$, $b$ and $c$, and the three initial conditions considered in this case, this region can be defined by $R_{h,3} = \{(\omega_p, \omega_q): 0 \leq \omega_p \leq 15, 0 \leq \omega_q \leq 5 \}$ in a non-exhaustive way. Unlike what happened within the stability region, these homogeneous states are not temporally stable, that is, they exhibit periodic behavior over time.

\subsubsection{Convergence outside the stability region of the local model}

Outside the stability region, the local model converges to a limit cycle generating periodic solutions. However, the inclusion of non-locality can cause, within a defined region of the parameter values and for particular initial conditions, the disappearance of the limit cycle, generating non-oscillating solutions with stable spatial patterns.

For the parameters considered for this case, $\omega_p$ and $\omega_q$ in the region defined by $\{ (\omega_p, \omega_q): 0 \leq \omega_p \leq 5, 10 \leq \omega_q \}$ we find  temporally stable spatial patterns for the initial conditions I.C.1 and I.C.2. When the initial condition I.C.3. is considered, the same results are obtained except for a few combinations of $\omega_p$ and $\omega_q$ inside of this region that produces spatio-temporal patterns. 

\begin{figure}[h]
    \centering
    \includegraphics[width=0.75\textwidth]{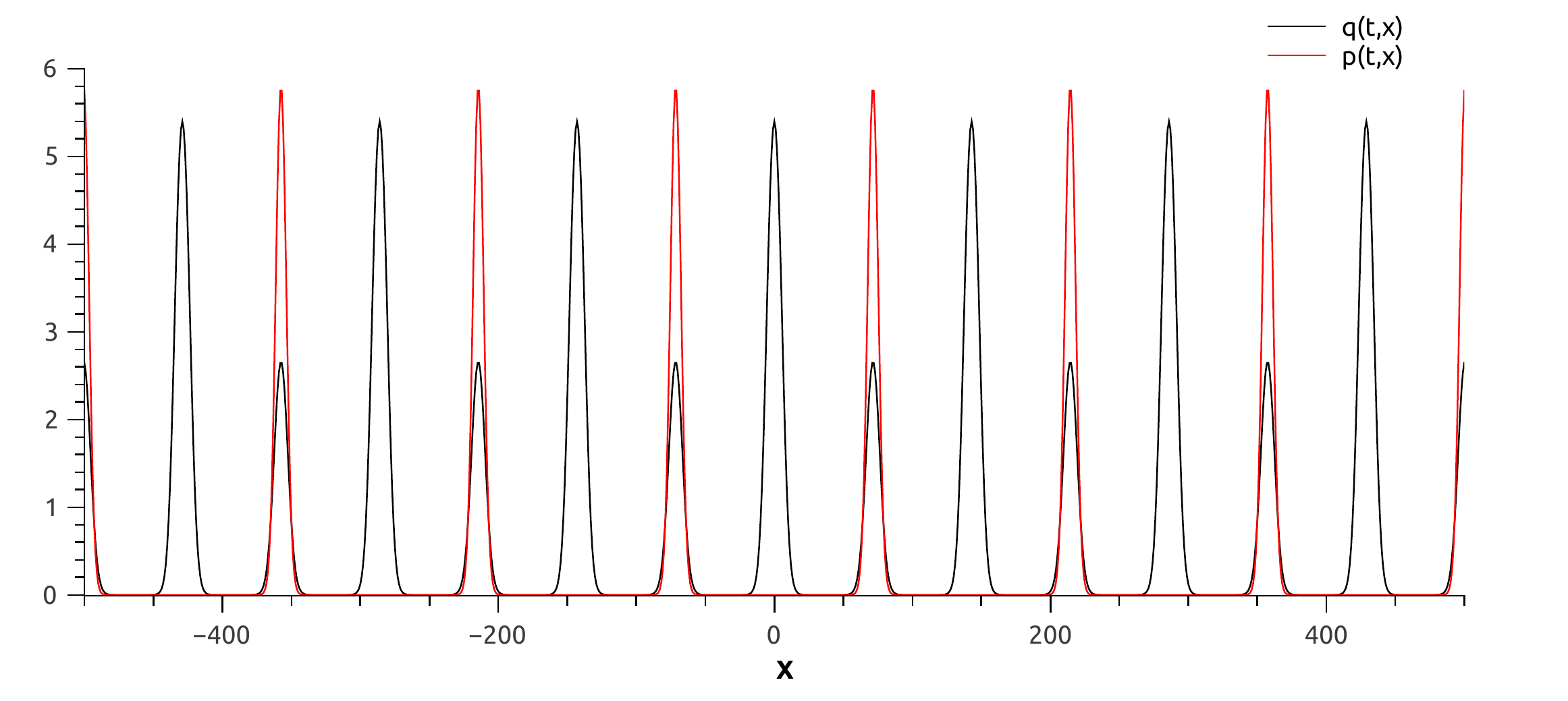}
    \caption{Temporally stable spatial patterns considering $a=0.8$, $b=0.05$, $c=0.25$, $\omega_p = 80$, $\omega_q=60$ and initial conditions I.C.1 or I.C.2.}
    \label{fig:stable-spatial-pattern}
\end{figure}

Other combinations of $\omega_p$ and $\omega_q$ producing temporally stable spatial patterns can be found outside this region, as for example $\omega_p = 80$ and $\omega_q=60$ that generates the same temporally stable spatial patterns of Fig. \ref{fig:stable-spatial-pattern}, for initial conditions I.C.1 and I.C.2. 

As mentioned in the previous cases, the counter-intuitive result, in this case showing  that there is convergence to a temporally stable spatial pattern, is not the most common one and only occurs for some combinations of $\omega_p$ and $\omega_q$.

\section{Discussion and Conclusions}

We want to  end this work by summarizing and analyzing our findings.
Here, we studied the dynamics of two interacting populations considering that the interaction can be non-local and of two different types: Competition and predation. Non-locality was included by means of an integral that evaluates the effect of the rest of the population within a defined radius of interaction $\omega_i$ where $i$ indicates the corresponding species. As the integral terms consist of a square kernel, all the individuals within the range are equally considered.  We solved both systems numerically and were able to observe that the inclusion of non-locality in these models allows for the emergence of a plethora of non-trivial behaviors which are not possible in the local problem and that are linked to the formation of spatial and spatio-temporal patterns. 
To highlight the new features due to the non-local terms we have chosen parameter values that correspond to different and well-studied behaviors in the local case.
Bearing in mind that diffusion is present in these equations, it is important to rule out the possibility of the emergence of patterns due to a Turing mechanism. In all cases, the choice of parameter values placed us far from  this scenario.

First, we should note that in the two types of interaction considered here, we were able to find a region in the plane $(\omega_u, \omega_v)$ for which the results for the non-local case were recovered. i.e. we observed that the solutions converge to the well-known homogeneous state, or eventually (in the predations case) to periodic oscillations. This is a known result for the case of a single population and was reported by \cite{fkk1,fkk2}. However determining the necessary and sufficient conditions for this to occur in the two-population models seems to be a non-trivial problem due to, at least, two factors. On the one hand, the asymptotic behavior of the solutions depends on the initial conditions, and on the other, the fact that it may be enough for a single population to present non-local interactions to obtain spatial patterns. While the previous results possess their own relevance, the most interesting phenomena correspond to the deviations from them.

In the competition model, when $a_{uv}<1$ and $a_{vu}<1$ (Sect. \ref{sec:comp_case_1}), we found steady spatial patterns always showing regions of local coexistence scattered with empty territories. That means that both populations survived and they partially or completely share the inhabited territories. This result is consistent with what happens in the local case, where for this choice of parameters, coexistence is allowed. We did not find patterns that could indicate global competitive exclusion, but we found that the spatial distribution of each species was strongly and differently affected by the size of the corresponding interaction radius. 

Completing the analysis of the results for the competition case, we considered  $a_{uv}>1$ and $a_{vu}>1$ (Sect. \ref{sec:comp_case_2}). Here we found situations of global competitive coexistence, but local exclusion, meaning that in the steady state there is a presence of both species though without sharing the habitat. This  situation is not possible when only local interactions are considered. In addition, we found that in terms of the values of the respective $\omega_v$ there is an alternation between global coexistence and global competitive exclusion. For example, if we screen the results for varying $\omega_v$ we find that for some values only $u$ survives, and for other values, it is $v$ which survives. In the middle of those regions of exclusion, in some cases we find global coexistence. This is an indication that we are not facing a real competitive coexistence but a situation where an otherwise displaced species takes advantage of the empty space left by the strongest one due to the non-local effects.

In addition to the results mentioned above, we found two other very interesting results. On the one hand what we call the phantom effect (when considering $a_{uv}<1$ and $a_{vu}>1$), that corresponds to a case where the final distribution of the survival species is affected by the initial presence and the corresponding $\omega$ range of the extinct species. This effect could be linked to the fact that the initial interaction between both species regulates the spatial modes that will survive and shape the patterned distribution of the survivor. The observed and already mentioned dependence of the steady state on the initial conditions is closely related to this in the sense that only the modes initially present will grow to define the spatial patterns.
On the other hand, we highlight the possibility of obtaining time-periodic solutions (Sect. \ref{sec:sp-temp-competition}). Our findings indicate that there is a region in the parameter space for which the solutions present oscillations, something impossible in the local case. To the best of our knowledge, this is a result that has not been previously reported.

Next, we will discuss our results for the predation system.
In this case, we considered two distinct cases well defined by Eq. (\ref{condT1}).
The knowledge derived from the local case tells us that the predator needs the presence of the prey to survive, while the opposite is not true. We determined the sufficient condition ($\omega_p < \omega_q$) for the spatially patterned survival region to be overlapped in the entire domain. In these cases, the abundance of the prey is lower than or equal to the abundance of the predator, the same happens, in general, with the height of the peaks. When the prey has a larger interaction range than the predator, the spatially patterned survival region may not overlap in the entire domain. In these cases, due to non-local effects, the predator can colonize regions where the prey is absent.

Unlike the competition system, this system can show oscillations in the local case due to the instability of the steady state through a Hopf bifurcation. 
Taking this feature into account, we explored the possibility of oscillations in the regions of parameter values where the equilibrium is steady and the oscillations are not present in the local case. One of the novel results presented in this work is precisely the emergence of oscillation in the corresponding range due to the non-local effects.
As a complementary result to the previous one, we have also found that the non-local effects can inhibit the oscillation that can be present in the local case under the proper conditions.

We hope that future research will analytically clarify the results presented by our numerical findings.

\section{Acknowledgements}

MIS is a postdoctoral fellow of CONICET.

\section{Supplementary Material}

\subsection*{Spatio-temporal patterns in competition model} \label{sec:sm-stpattern}

Each $\omega_v$-level of Fig. \ref{fig:STP-parameters}. 

\begin{figure}[h]
     \centering
     \begin{subfigure}[b]{0.49\textwidth}
         \centering
         \includegraphics[width=\textwidth]{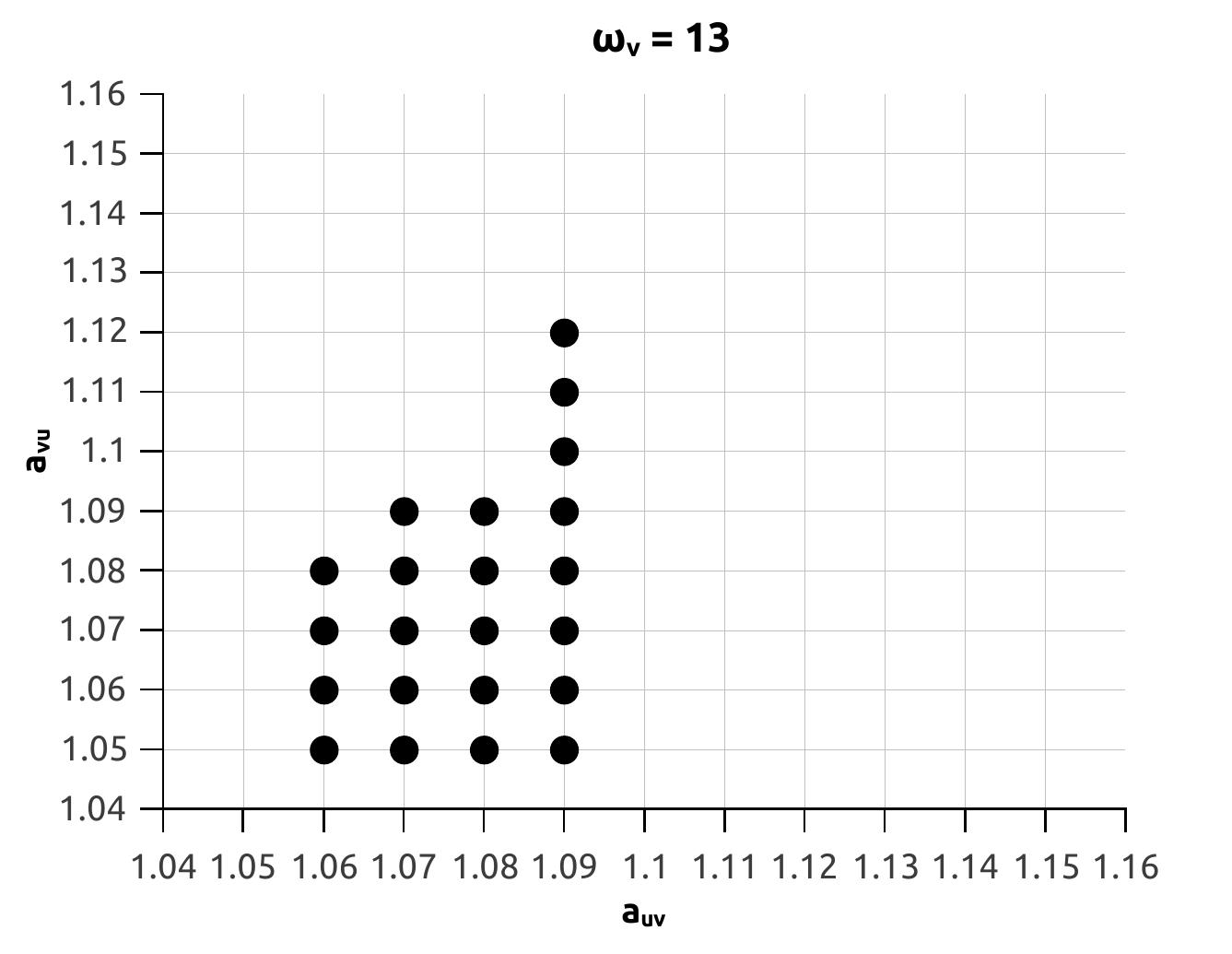}
     \end{subfigure}
     \hfill
     \begin{subfigure}[b]{0.49\textwidth}
         \centering
         \includegraphics[width=\textwidth]{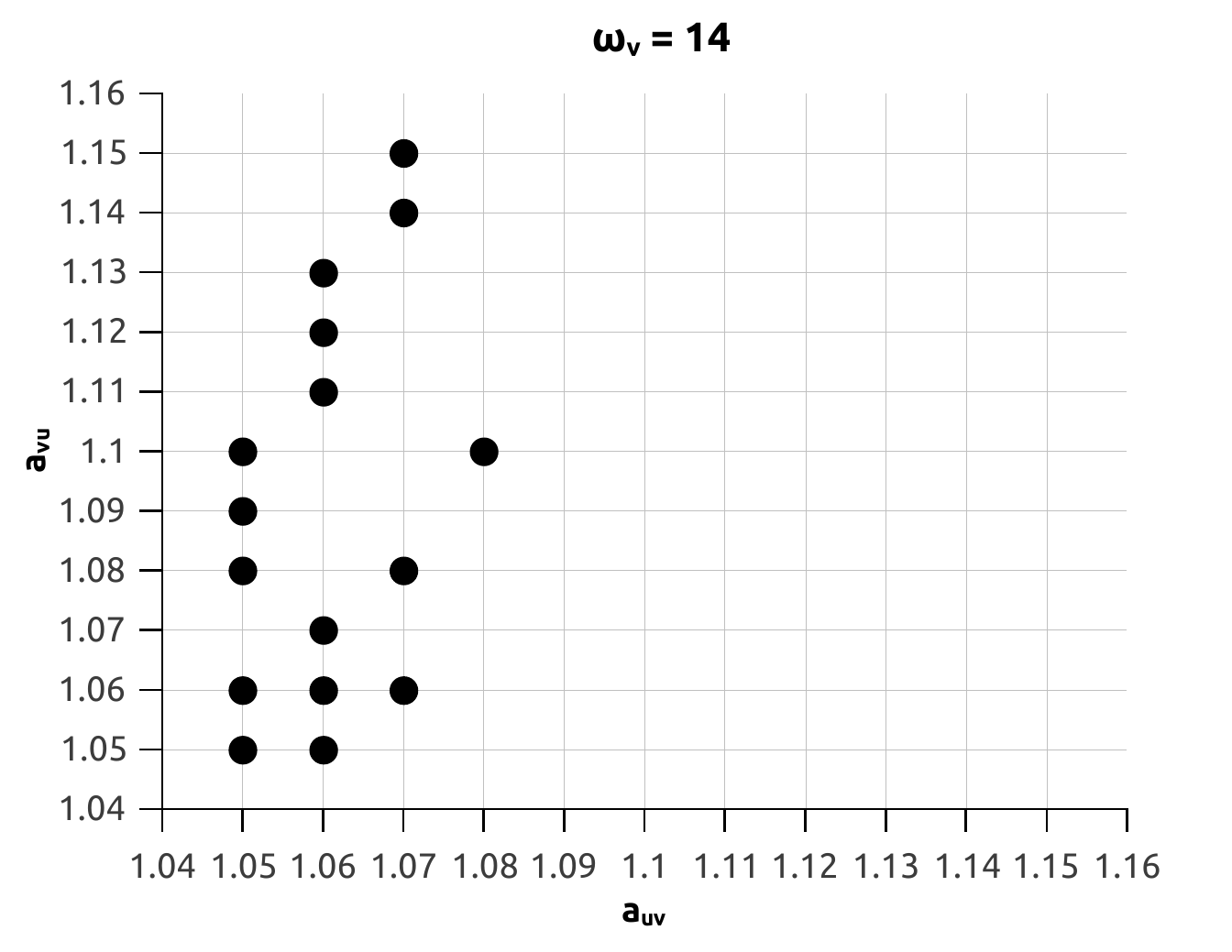}
     \end{subfigure}
     \begin{subfigure}[b]{0.49\textwidth}
         \centering
         \includegraphics[width=\textwidth]{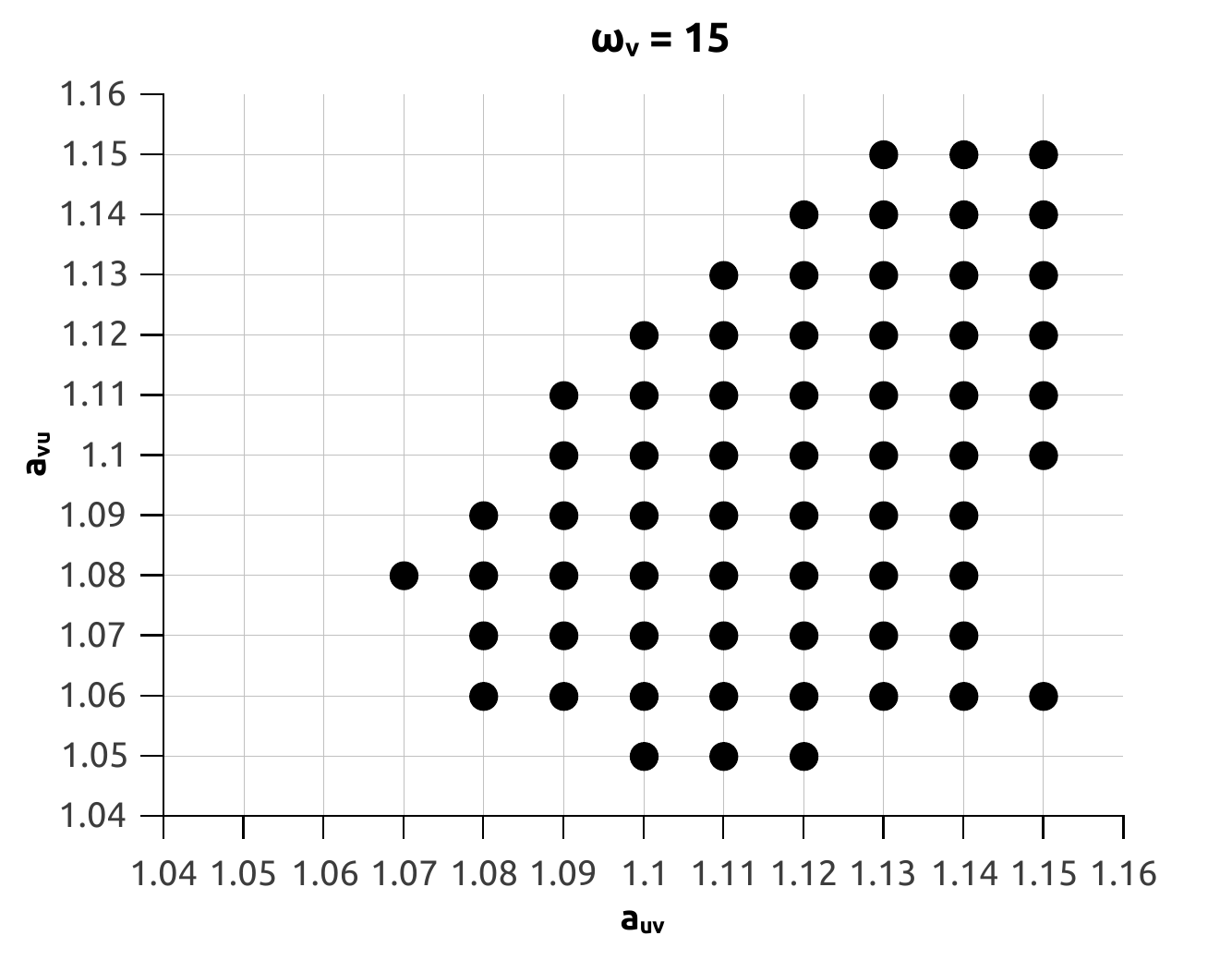}
     \end{subfigure}
     \hfill
     \begin{subfigure}[b]{0.49\textwidth}
         \centering
         \includegraphics[width=\textwidth]{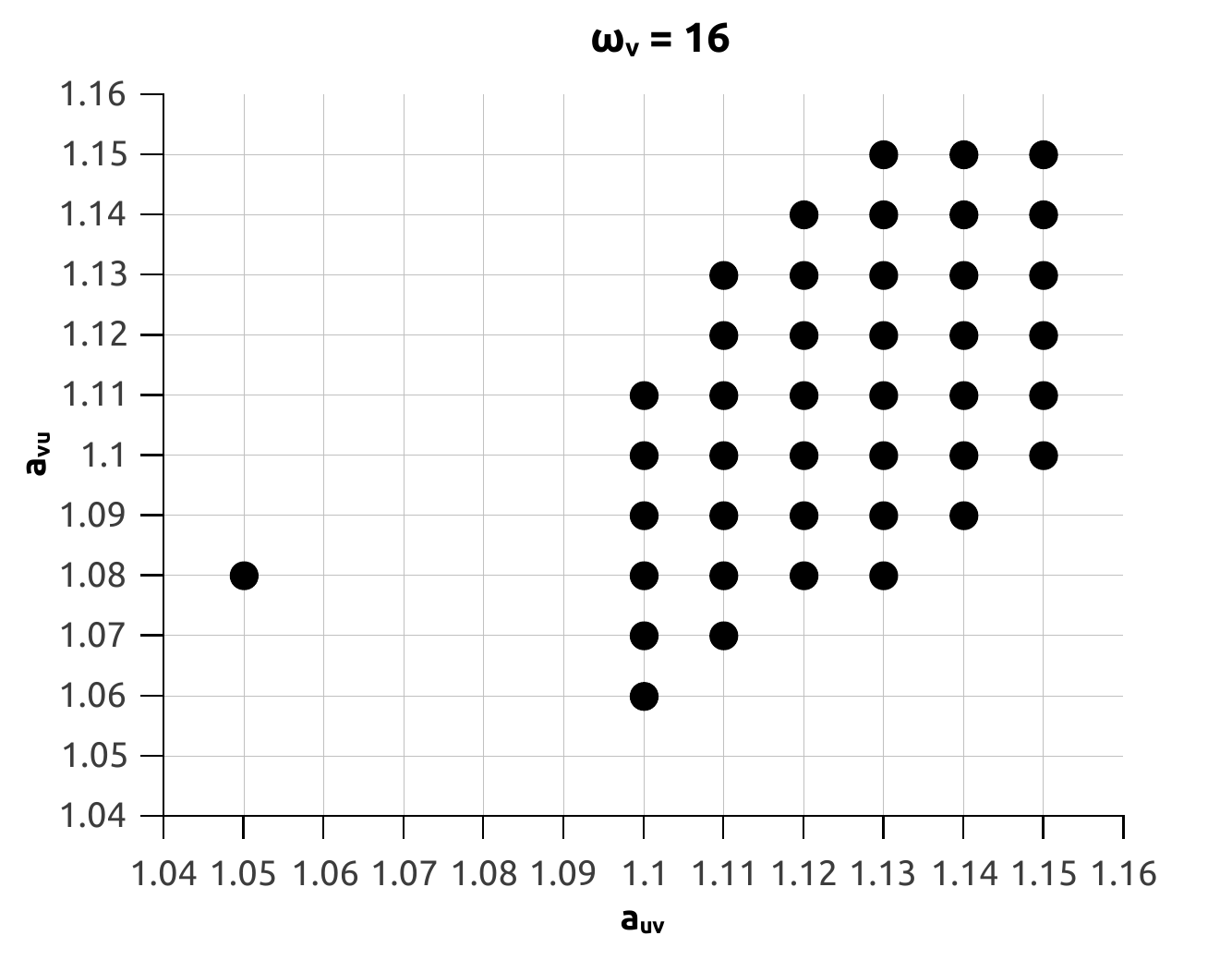}
     \end{subfigure}
      \caption{$\omega_v$-levels of Fig. \ref{fig:STP-parameters}}
        \label{fig:three graphs}
\end{figure}
\begin{figure} \ContinuedFloat
     \centering 
     \begin{subfigure}[b]{0.49\textwidth}
         \centering
         \includegraphics[width=\textwidth]{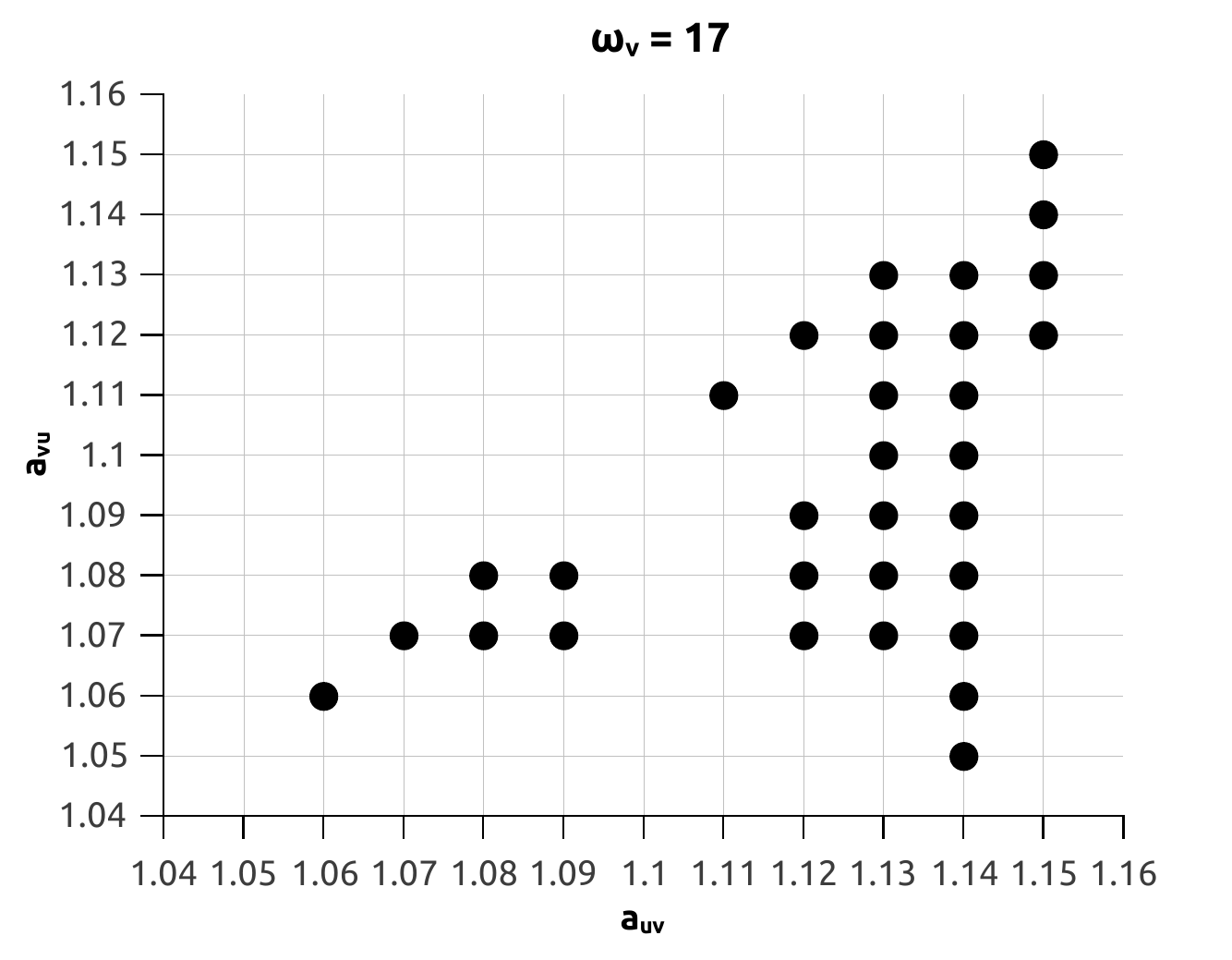}
     \end{subfigure}
     \hfill
     \begin{subfigure}[b]{0.49\textwidth}
         \centering
         \includegraphics[width=\textwidth]{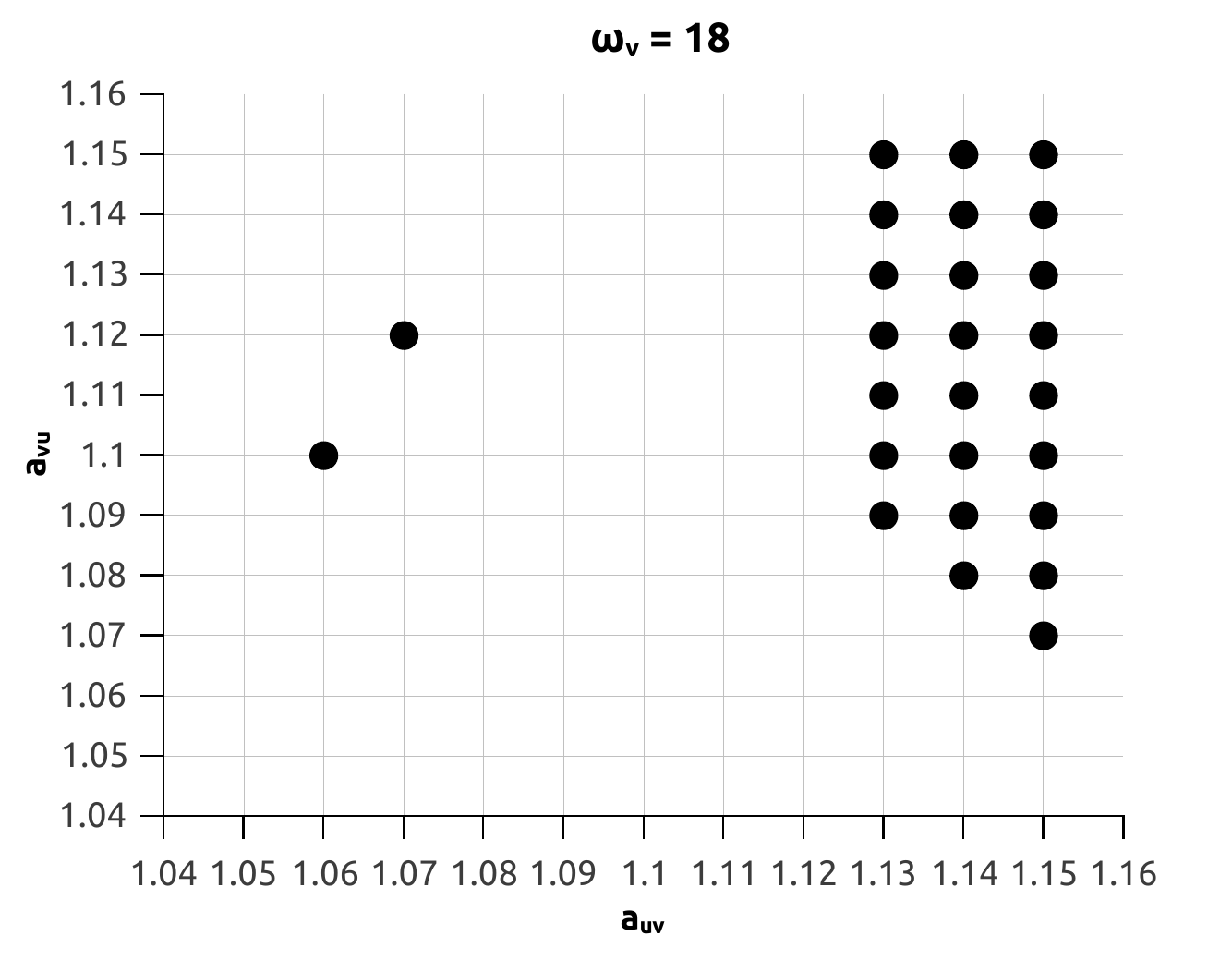}
     \end{subfigure}
     \hfill
     \begin{subfigure}[b]{0.49\textwidth}
         \centering
         \includegraphics[width=\textwidth]{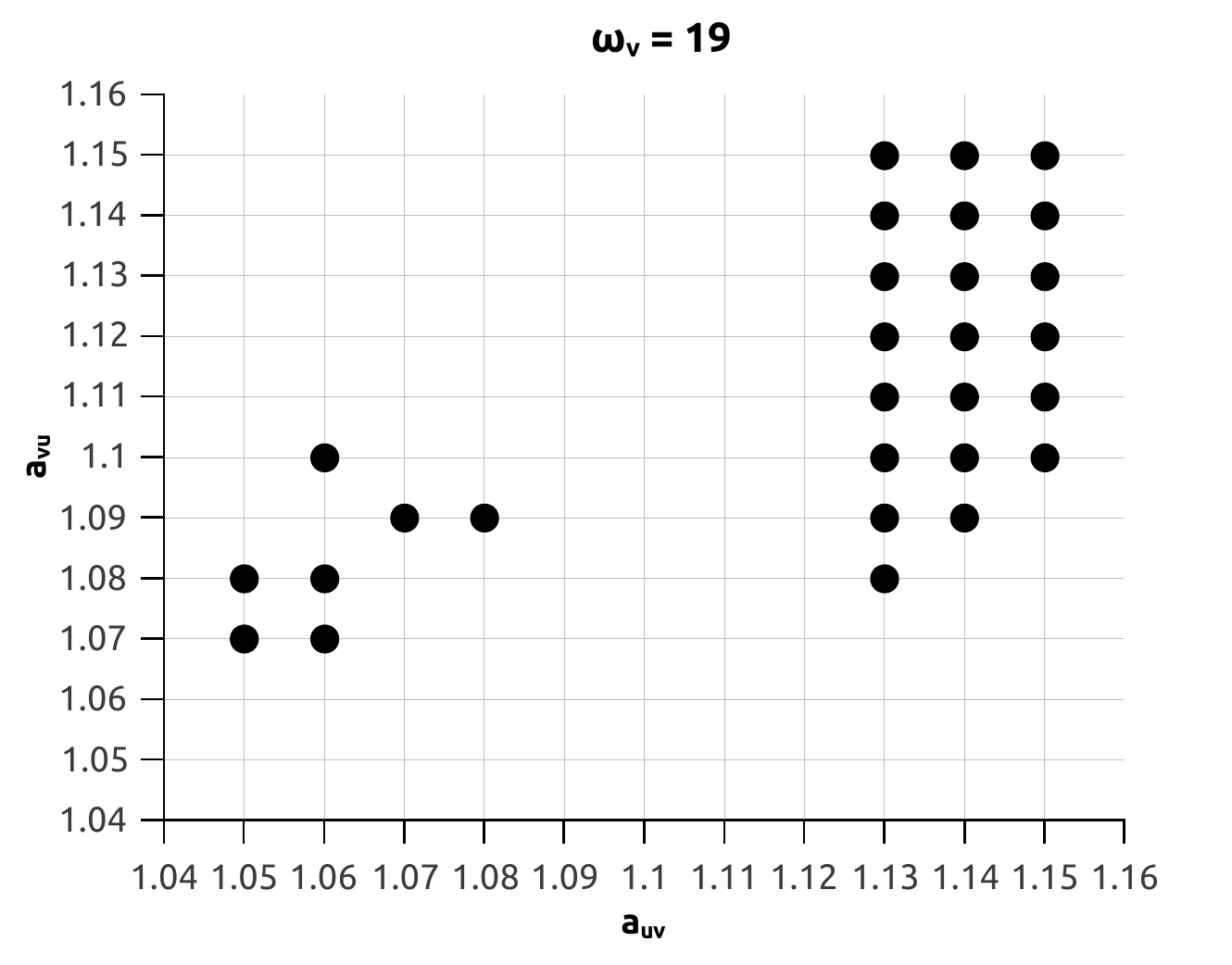}
     \end{subfigure}
     \hfill
     \begin{subfigure}[b]{0.49\textwidth}
         \centering
         \includegraphics[width=\textwidth]{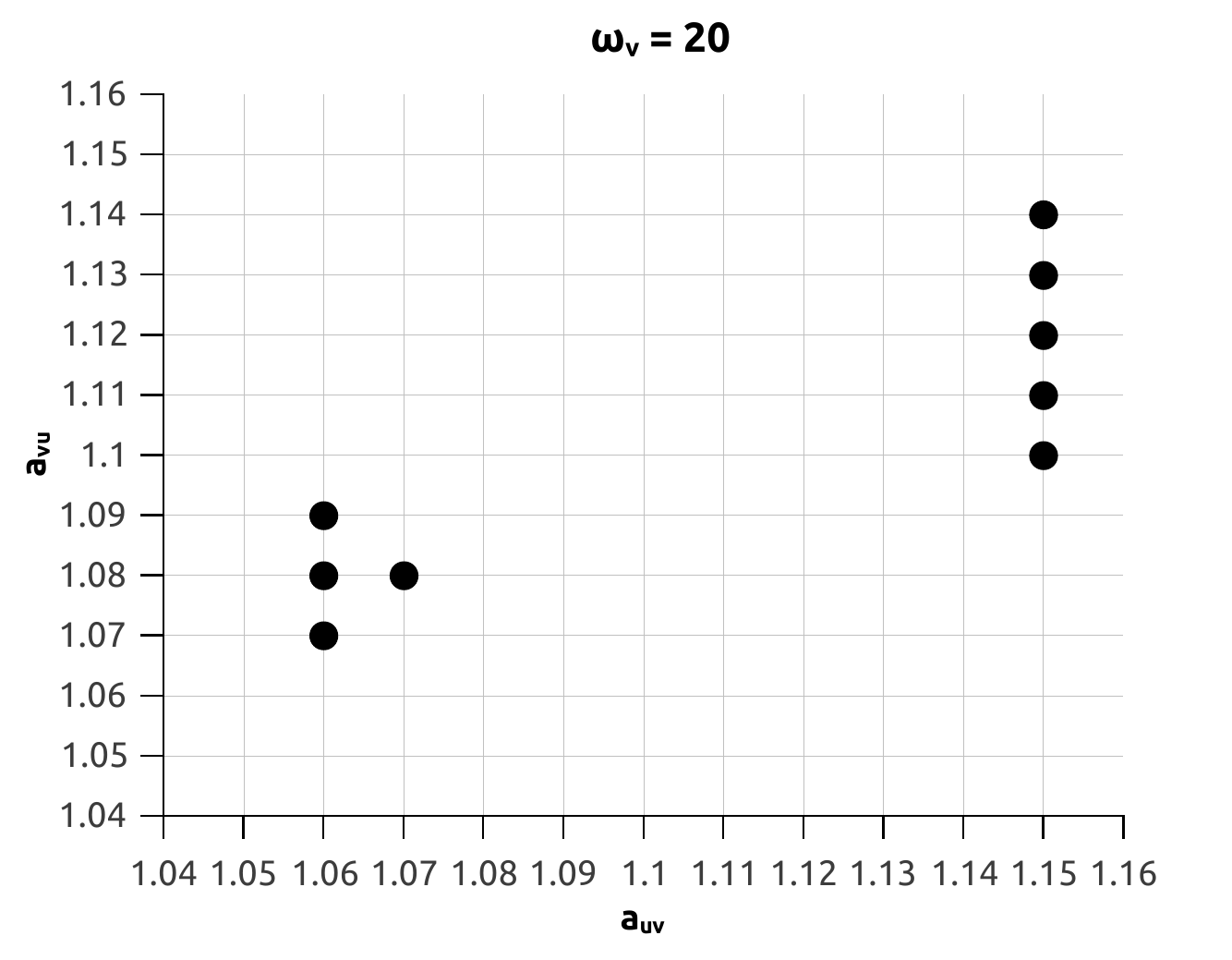}
     \end{subfigure}
        \caption{$\omega_v$-levels of Fig. \ref{fig:STP-parameters} (cont.)}
        \label{fig:three graphs}
\end{figure}

\end{document}